\documentclass[10pt]{article}

\usepackage[utf8]{inputenc}
\usepackage[english]{babel}
\usepackage[toc,page]{appendix}

\usepackage{amsmath,amsfonts,amssymb,amsthm}
\usepackage{graphicx}
\usepackage{float}
\usepackage{hyperref}
\usepackage{dsfont}
\usepackage{subcaption}
\usepackage[format=plain,font=small,textfont=it]{caption}
\usepackage[dvipsnames]{xcolor}
\usepackage{multirow} 
\usepackage{bookmark}
\usepackage{mathrsfs}

\graphicspath{{img/}}

\newcommand{\ket}[1]{|#1\rangle}
\newcommand{\bra}[1]{\langle #1|}
\newcommand{\sket}[1]{|#1]}
\newcommand{\sbra}[1]{[ #1|}

\newcommand{\braket}[2]{\langle #1 \vert #2 \rangle}
\newcommand{\sbraket}[2]{[ #1 \vert #2 \rangle}
\newcommand{\brasket}[2]{\langle #1 \vert #2 ]}
\newcommand{\sbrasket}[2]{[ #1 \vert #2 ]}

\newcommand{\R}{\mathbb{R}}
\newcommand{\C}{\mathbb{C}}

\renewcommand{\Re}{\mathrm{Re}}
\renewcommand{\Im}{\mathrm{Im}}

\renewcommand{\d}{\mathrm{d}}
\begin{document}

\title{Geometry from local flatness in Lorentzian spin foam theories}

\author{\Large{Pietro Don\`a\footnote{dona.pietro@gmail.com},}
\smallskip \\ 
\small{\textit{Center for Space, Time and the Quantum, 13288 Marseille, France}}\\
\small{\textit{Department of Physics and Astronomy, University of Western Ontario, London, ON N6A 5B7, Canada}}
}

\date{\today}

\maketitle

\begin{abstract}
Local flatness is a property shared by all the spin foam models. It ensures that the theory's fundamental building blocks are flat by requiring locally trivial parallel transport. In the context of simplicial Lorentzian spin foam theory, we show that local flatness is the main responsible for the emergence of geometry independently of the details of the spin foam model. We discuss the asymptotic analysis of the EPRL spin foam amplitudes in the large quantum number regime, highlighting the interplay with local flatness.
\end{abstract}

%%%%%%%%%%%%%%%%%%%%%%%%%%%%%%%%%%%%%%%%%%%%%%%%%%%%%%%%%%%%%%%%%%%%%%%%%%%%%%%%%%%%%%%%%%

\section{Introduction and motivations}
\label{sec:intro}

%%%%%%%%%%%%%%%%%%%%%%%%%%%%%%%%%%%%%%%%%%%%%%%%%%%%%%%%%%%%%%%%%%%%%%%%%%%%%%%%%%%%%%%%%%

Spin foam theory is a promising candidate for quantization of gravity and is often referred to as the covariant formulation of Loop Quantum Gravity. It attempts to define a Lorentzian background-independent path integral for General Relativity regularized on a fixed triangulation of a space-time manifold. The theory assigns quantum numbers representing geometric quantities to the 2-complex dual to the triangulation. The path integral sums over all the possible quantum numbers describe a sum over quantum geometries. Moreover, spin foam theory assigns transition amplitudes to states in the kinematical Hilbert space of Loop Quantum Gravity associated with the triangulation boundary. 

\medskip

The EPRL-FK model \cite{Engle:2007wy,Freidel:2007py} is the most promising spin foam theory available that is currently being developed (see \cite{Rovelli:2014ssa, Perez:2012wv} for a pedagogical introduction). Its most celebrated and successful result is the emergence of Regge geometries and the Regge action in the large quantum numbers regime of the spin foam amplitudes \cite{Barrett:2009mw}. There are many extensions and applications including but not limited to the Euclidean EPRL model \cite{Barrett:2009gg}, the topological $SU(2)$ BF theory \cite{Barrett:2009as,Dona:2017dvf}, to general cellular decomposition known as the KKL model \cite{Kaminski:2009fm, Dona:2020yao}, to generic $SL(2,\C)$ tensors \cite{Dona:2020xzv}, and have a numerical confirmation \cite{Dona:2019dkf,Gozzini:2021kbt}. The original and the majority of the related works are based on the bivector reconstruction theorem \cite{Barrett:1997gw} a map from Euclidean or Lorentzian 4-simplices to a collection of bivectors satisfying (linear simplicity) constraints. This elegant theorem was instrumental in driving the geometrical intuition of spin foam models, and Regge calculus is the best tool to study the semiclassical limit \cite{Immirzi:1996dr,Barrett:1998gs}.

\medskip

The EPRL spin foam model, and all of its extensions, have very complicated recipes with many components mixed. The traditional semiclassical analysis of the EPRL spin foam model does not disentangle the role played by each ingredient and blends all of them. A clear picture of each contribution is fundamental to proposing improvements, extensions, and generalizations of the model and increasing our understanding of the theory. We separate the analysis into two main components: the \emph{local flatness} of the spin foam vertices and the role of \emph{saddle point} equations.

We abandon the bivector reconstruction theorem favoring gauge covariant geometrical objects like framed polyhedra, dihedral angles, and twist angles. This description follows, and is inspired by, previous work on discrete holonomy-flux geometries \cite{Dittrich:2008ar} and on twisted geometries \cite{Freidel:2010aq,Freidel:2013fia, Speziale:2012nu,Anza:2014tea}. We parametrize the wedge holonomies using a set of four framed planes for each edge and a complex angle to each wedge.

\medskip

Spin foam models are built with \emph{locally flat} triangulations. Any parallel transport in a 4-simplex is trivial. This property is very general, and all the spin foam models in the literature share it. When we impose \emph{local flatness} to a set of general holonomies, we find that geometry naturally emerges. The holonomies are Regge-like. Since local flatness does not involve areas, we cannot expect to reconstruct a full Regge geometry. The holonomies contain information on just the angles of the emerging geometry. Suppose we also provide a closure condition for each edge. In that case, we can identify framed tetrahedra and a unique Regge geometry emerge up to a scale factor (if we ignore the topological sector and discrete symmetries). This result is entirely independent of the details of the model itself. Therefore we expect the emergence of geometry to be a typical result of all locally flat spin foam models.

We study the large quantum number regime of the EPRL spin foam amplitude with \emph{saddle point} techniques. The resulting equations provide a closure equation to each edge and align the boundary framed planes with the ones parametrizing the wedge holonomies. As a result, at the saddle point, the amplitude assumes a simple form in terms of the complex angles of the wedge holonomies mixing boost and rotation angles with the Immirzi parameter. Only if local flatness is considered the saddle point action assumes the form of the Regge action completing the connection with discrete General Relativity.

\medskip

As a coproduct of our study, we provide an explicit form of the action at the saddle point in terms of gauge invariant quantities. In particular, its imaginary part was not studied enough in the literature (because it is usually deemed as irrelevant \cite{Barrett:2009mw} or gauge-fixed \cite{Dona:2019dkf}). However, having complete control of all the phases is necessary for the numerical evaluation of the amplitude \cite{Dona:2019dkf, Asante:2022lnp}. 

We complete our analysis by also studying the case of extended triangulations with many vertices. First, we consider them fixed and large instead of summing over the bulk quantum numbers. The saddle point equations reduce to a closure condition at every edge and gluing equations that tell us how to glue the geometry of the various vertices together. We complement the analysis by looking at the role of the sum over bulk degrees of freedom. We follow the seminal work of \cite{Hellmann:2013gva} and find that the face holonomy is constrained to be a 4-screw with proportional boost and rotation angles. The naive flatness problem arises immediately if this condition is supplemented by local flatness and gluing conditions.

%%%%%%%%%%%%%%%%%%%%%%%%%%%%%%%%%%%%%%%%%%%%%%%%%%%%%%%%%%%%%%%%%%%%%%%%%%%%%%%%%%%%%%%%%%

\section{Spinors, framed planes and maps between them}
\label{sec:spinors}

%%%%%%%%%%%%%%%%%%%%%%%%%%%%%%%%%%%%%%%%%%%%%%%%%%%%%%%%%%%%%%%%%%%%%%%%%%%%%%%%%%%%%%%%%%

Spinors are elements of the vector space $\C^2$. We indicate a spinor and its conjugate transpose with $\ket{z}$ and $\bra{z}$.
\begin{equation}
\ket{z} := \left(\begin{array}{c}
z_{0}\\
z_{1}
\end{array}\right)
\ ,
\qquad 
\bra{z} := \left(
\bar{z}_{0},\  \bar{z}_{1}
\right)
\ .
\end{equation}
We use the Dirac notation to simplify the bookkeeping of spinorial indices. The spinorial space comes with a natural inner product given by $\braket{w}{z} := \bar{w}_0 z_0 + \bar{w}_1 z_1$ and a duality map $\mathcal{J} : \C^2 \to \C^2$ 
\begin{equation}
	\mathcal{J} \ket{z} = \sket{z} := \left(\begin{array}{c}
-\bar{z}_{1}\\
\bar{z}_{0}
\end{array}\right)\ .
\end{equation}
Moreover, a spinor and its dual form an orthogonal basis of $\C^2$. Spinors also have many interesting properties useful for performing calculations. We report some of them in Appendix~\ref{app:aspinor}.

\medskip

\begin{figure}[H]
	\centering
	\includegraphics[scale=0.5]{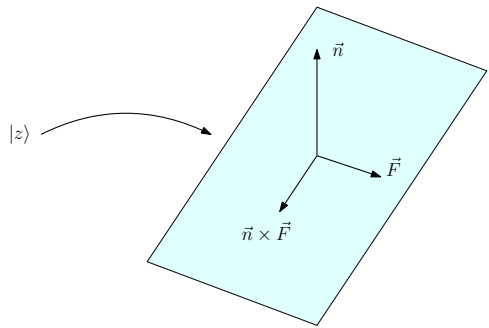}
	\caption{\label{fig:framedplane} Pictorial representation of the reference frame associated to a spinor $\ket{z}$. The magnitude of the spinor determines the plane and the phase of the spinor the frame of the plane.} 
\end{figure}

A spinor naturally defines also a basis of $\R^3$ given by $(\vec{n}, \vec{F}, \vec{n}\times \vec{F})$ built from the matrix elements of the Pauli matrices
\begin{equation}
\label{eq:frame}
\bra{z}\vec{\sigma} \ket{z} = -\vec n \ , \qquad \text{and} \qquad \sbra{z}\vec{\sigma}\ket{z} = i \vec{F}+\vec{n}\times \vec{F} \ .
\end{equation}
In the following, we will work with unit norm spinors $\braket{z}{z}=1$ to simplify the formulas if not specified otherwise. In this case, the associated $\R^3$ basis is orthonormal. Each spinor identifies a (framed) plane in $\R^3$ orthogonal to $\vec{n}$ equipped with a frame given by the vector $\vec{F}$ (sometimes we will call it frame vector \cite{Freidel:2013fia}). See Figure~\ref{fig:framedplane} for a pictorial representation. Intuitively, the magnitude of the spinor determines the vector $\vec{n}$ while its phase (flag) characterizes completely the frame vector $\vec{F}$. 

This geometrical interpretation is the same one used in the geometrical interpretation of loop quantum gravity states given by \emph{twisted geometries} \cite{Freidel:2010aq,Rovelli:2010km,Freidel:2010bw,Dupuis:2012yw}.

\medskip

A general $SL(2,\C)$ group element $g$ can be parametrized by two (unit) spinors $\ket{z}$, $\ket{w}$ and a complex number $\omega$
\begin{equation}
\label{eq:gdiagonal}
g = e^{\frac{\omega}{2}} \ket{w}\bra{z}+ e^{-\frac{\omega}{2}} \sket{w}\sbra{z} \ .
\end{equation}
The group element $g$ maps the source framed plane $\ket{z}$ into the target one $\ket{w}$. The complex angle $\omega$ encodes the boost angle and part of the rotation between the planes (see Appendix~\ref{app:diag} for more details). 
The parametrization \eqref{eq:gdiagonal} is very redundant. Three real numbers parametrize each unit spinor (i.e., two angles to determine $\vec{n}$ and a third to determine $\vec{F}$ in the orthogonal plane). The complex number $\omega$ contributes to the count with two real degrees of freedom. Eight real parameters are too many to specify a $SL(2,\C)$ group element. However, a direct calculation shows that only one between the phases of the two spinors and $\Im\omega$ is independent. Shifting any of those reabsorbs any shifts of the others. Notice, that also the dual group element $(g^{-1})^\dagger$ defines a map between the source and target frames $\ket{z}$, $\ket{w}$ with opposite chirality and can be obtained from \eqref{eq:gdiagonal} sending $\omega \to -\omega^*$.

\medskip

The parametrization \eqref{eq:gdiagonal} is analog to the parametrization of the $SU(2)$ holonomy of the Ashtekar connection in \emph{twisted geometries} \cite{Freidel:2010aq,Rovelli:2010km,Freidel:2010bw,Dupuis:2012yw} and can be recovered from \eqref{eq:gdiagonal} setting $\Re \omega = 0$ and removing the redundant $\Im \omega$. A form equivalent to \eqref{eq:gdiagonal} is derived in the context of covariant twisted geometries in terms of twistors in \cite{Speziale:2012nu}.

\medskip

%The $SL(2,\C)$ action \eqref{eq:gaction} on the spinorial space is natural and represent a map between the framed planes $\ket{z}$ and $\ket{w}$. 
While the spinorial space carries a representation (the fundamental one) of $SL(2,\C)$, the three dimensional space $\R^3$ spanned by $(\vec{n}, \vec{F}, \vec{n}\times \vec{F})$ does not. If we want to interpret the action of the Lorentz group on $\R^3$ we have to embed it in a larger space where a representation is realized. The embedding is not unique, and the interpretation depends on this choice. Among many possibilities, the EPRL model and the study of general $SL(2,\C)$ invariants\cite{Dona:2020xzv} suggest one in terms of $\gamma$-simple bivectors. 
We represent a bivector with electric and magnetic part $\vec{E}$ and $\vec{B}$ with a complex vector given by $\Pi = \vec{E}+i\vec{B}$. In electromagnetism, it is known as the Riemann-Silberstein vector. If the electric and magnetic parts of the bivector are proportional $\vec{B}= \gamma \vec{E}$ we say the bivector is $\gamma$-simple (in the canonical frame). In this case we parametrize it with a complex number $j + i \gamma j$ and a unit vector $\vec{n}$. Or, in terms of the spinor $\ket{z}$,
\begin{equation}
\label{eq:Pi}
\Pi = j(1+i\gamma) \vec{n} = -j(1+i\gamma)  \bra{z} \vec{\sigma} \ket{z} \ .
\end{equation}
In the context of the EPRL spin foam model, we interpret $j$ as quanta of areas and $\gamma$ as the Immirzi parameter. This particular bivector is the consequence of constructing and implementing the linear simplicity constraints in the EPRL model. 
%Notice, this is a very specific self-dual bivector. In particular if we consider the representation of $\Pi$ as a rank-2 tensor, then $B = (1-\gamma \star) \Pi$ is simple and $B \propto \star t \wedge n$ where $t=(1,0,0,0)$ is the time-like canonical direction and $n = (0,\vec{n})$ is the trivial embedding $\vec{n}$. We can think of \eqref{eq:Pi} as a self-dual $\gamma$-simple bivector.
The bivector $\Pi$ transform in the finite-dimensional representation of $SL(2,\C)$ usually denoted as $(0,1)$. Therefore, we can see a group element \eqref{eq:gdiagonal} as a map between the two $\gamma$-simple bivectors associated with $\ket{z}$ and $\ket{w}$ \footnote{Note that if we apply $g$ to a $\gamma$-simple bivector, that is not the one associated to the source spinor of $g$ the transformation is complicated and, in general, it will not be mapped into another $\gamma$-simple bivector.}
\begin{equation}
\label{eq:bivector}
g\triangleright \Pi = j(1+i\gamma) \bra{z} g^{-1} \vec{\sigma} g \ket{z} = e^{\frac{-\omega+\omega}{2}} j(1+i\gamma) \bra{w} \vec{\sigma}  \ket{w} = \Pi' \ . 
\end{equation}
Since $\Pi$ transforms in the $(0,1)$ representation, we have to use fundamental representation and its dual. To make the bilinear \eqref{eq:Pi} transform in the $(0,1)$ representation we need both $\ket{z}$ and $\bra{z}$ with the same chirality. This is the reason for the presence of the inverse group element \eqref{eq:bivector}. 
%We compute the transformation of the frame of the plane $\ket{z}$ using its definition the definition 
%\begin{equation}
%i \vec{F}'+\vec{n}'\times \vec{F}' = e^{-i\Im \omega}\sbra{z} \vec{\sigma} \ket{z} =i \left( \cos\Im\omega \,\vec{F}- \sin\Im\omega \, \vec{n}\times \vec{F} \right) + \left( \cos\Im\omega \, \vec{n}\times \vec{F} + \sin\Im\omega \, \vec{F} \right) \ .
%\end{equation}
%The holonomy $g$ acts on the frame of the plane $\ket{z}$ as a rotation of an angle of $\Im \omega$.

\medskip

Since we are interested in keeping the results as general as possible, we would not rely on this interpretation, so we will not specify any embedding until we talk about the EPRL model. We consider \eqref{eq:gdiagonal} as a map between spinors that only then do we interpret as framed planes. This is what we mean, with a slight abuse of language, when we will say that \eqref{eq:gdiagonal} is a map between framed planes.

\medskip

\begin{figure}[H]
	\centering
	\includegraphics[scale=0.5]{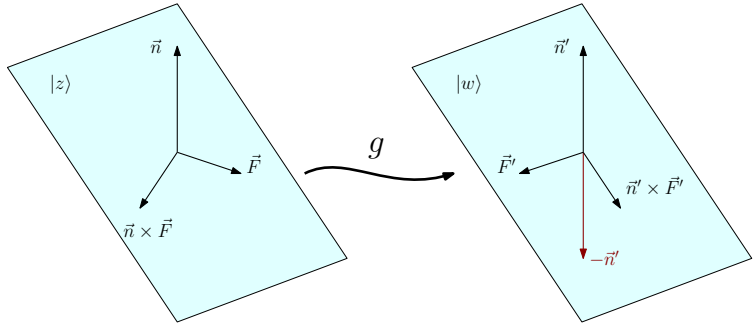}
	\caption{\label{fig:holonomy} Pictorial representation of the holonomy as a map between reference frames.} 
\end{figure}

In the following, we shift the focus from general framed planes to planes representing the faces of framed polyhedra. With this goal in mind, it is convenient to use a holonomy that maps the framed plane $\ket{z}$ into the framed plane dual to $\ket{w}$
\begin{equation}
\label{eq:gdiagonal-oriented}
g = e^{\frac{\omega}{2}} \sket{w}\bra{z} - e^{-\frac{\omega}{2}} \ket{w}\sbra{z} \ .
\end{equation}
In this way, the framed planes associated with the faces of polyhedra will have all outgoing normals. This is just a change of convention to ease the future geometrical interpretation. See Figure~\ref{fig:holonomy} for a pictorial representation. 

%%%%%%%%%%%%%%%%%%%%%%%%%%%%%%%%%%%%%%%%%%%%%%%%%%%%%%%%%%%%%%%%%%%%%%%%%%%%%%%%%%%%%%%%%%

\section{Wedge holonomies and local flatness}
\label{sec:localflatness}

%%%%%%%%%%%%%%%%%%%%%%%%%%%%%%%%%%%%%%%%%%%%%%%%%%%%%%%%%%%%%%%%%%%%%%%%%%%%%%%%%%%%%%%%%%

\begin{figure}[H]
\centering
\includegraphics[scale=0.7]{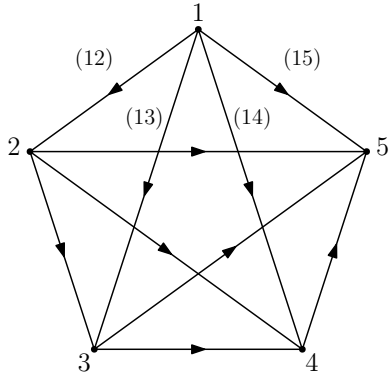}
\caption{\label{fig:4-simplex} The 2-complex of a 4-simplex. The edges are numbered from $1$ to $5$. A couple of edges label the wedges (faces in a vertex). To keep the picture clean, we just explicitly wrote the name of the wedges involving the edge $1$. We oriented the wedges $ab$ with $a<b$ such that $a$ is the source and $b$ is the target of the wedge.}
\end{figure}

We consider the 2-complex of a 4-simplex as illustrated in Figure~\ref{fig:4-simplex}. It comprises five edges (dual to tetrahedra) and ten faces (dual to triangles). A face in a vertex is also called a wedge and contains two edges. We orient the wedges indicating the source and the target edge. We label each edge with an index $a=1,\ldots,5$ and each wedge with the couple $ab$ where $a$ is the source, and $b$ is the target.

We associate a $SL(2,\C)$ group element $g_{ab}$ to every wedge $(ab)$ representing the holonomy along that wedge. It describes the parallel transport from edge $a$ to edge $b$. Following \eqref{eq:gdiagonal-oriented}, we parametrize it as
\begin{equation}
\label{eq:wedgeholo}
g_{ab} = e^{\frac{\omega_{ab}}{2}} \sket{z_{ba}}\bra{z_{ab}} - e^{-\frac{\omega_{ab}}{2}} \ket{z_{ba}}\sbra{z_{ab}} \ .
\end{equation}
The spinor $\ket{z_{ab}}$ identifies a framed plane at the source edge $a$ and the spinor $\ket{z_{ba}}$ identifies a framed plane at the target edge $b$. Given an edge $a$, four holonomies involve it, and four framed planes, identified by the spinors $\ket{z_{ab}}$ with $b\neq a$, are associated with it. With this notation, the inverse of \eqref{eq:wedgeholo} and the holonomy of the wedge with inverted orientation are related by $g_{ba} = - g_{ab}^{-1}$.
%As a reminder, with our conventions the framed planes are always identified by the spinors $\ket{z_{ab}}$ 

\medskip
Spin foam models are \emph{locally flat}: they are built using flat 4-simplices (or, more general, four-dimensional cells). In terms of holonomies, a 4-simplex is flat if the parallel transport along any cycle is trivial \footnote{A cycle is a closed, ordered collection of wedges. For example $(abc)= \{(ab), (bc), (ca)\}$ is a 3-cycle. A cycle made of three wedges.}. For example, for a 3-cycle $(abc)$, the parallel transport is trivial if
\begin{equation}
\label{eq:3cycleflatness}
g_{ca}g_{bc}g_{ab} =  \mathds{1} \ .
\end{equation}
For a simplicial vertex (as Figure~\ref{fig:4-simplex}), imposing \eqref{eq:3cycleflatness} on all the 3-cycles is sufficient to guarantee that the parallel transport on any cycle is trivial\footnote{All the cycles of a general graph are generated composing a the \emph{fundamental cycles}. For a 4-simplex, the fundamental cycles are given by a set of six independent 3-cycles. We will consider all 3-cycles even if redundant to deal with symmetric equations.}. The generalization, to a general cell \cite{Kaminski:2009fm, Ding:2010fw, Dona:2020yao} is straightforward. However, we need to be careful to consider the correct number and order of cycles (higher-order cycles are probably necessary \cite{Bahr:2018vvq, Dona:2020yao}).

\medskip

A general set of holonomies is not \emph{locally flat}. We impose it as a requirement. We find some constraints on the spinors and complex angles parametrizing the wedge holonomies \eqref{eq:wedgeholo}. After some lengthy but straightforward algebra (that we report in detail in Appendix~\ref{app:derivation_spherical}), we find that the equations \eqref{eq:3cycleflatness} on every 3-cycle are equivalent to
\begin{equation}
\label{eq:cosine_equations}
\cosh (\omega_{ab}+i\xi_{ab}^{c}) =\cos\hat\theta_{ab}^{c} \ ,
\end{equation}
for each wedge $ab$ and
\begin{equation}
\label{eq:sine_equations}
\sin\phi_{ac}^{b}\sinh(\omega_{ab}+i\xi_{ab}^{c})	= \sin\phi_{ab}^{c}\sinh(\omega_{ca}+i\xi_{ac}^{b}) \ .
\end{equation}
for each couple of wedges $ab$ and $ac$. The angles $\hat\theta_{ab}^{c}$ and $\xi_{ab}^{c}$ are functions of the spinors and have an interesting geometrical interpretation. 

\medskip

The angle $\hat\theta_{ab}^{c}$ is the \emph{generalized dihedral angle} and is given by the expression
\begin{equation}
\label{eq:spherical_cosine_law}
\cos\hat\theta_{ab}^{c} = 
\frac{-|\brasket{z_{ca}}{z_{cb}}|^{2}+|\braket{z_{ab}}{z_{ac}}|^{2}|\braket{z_{ba}}{z_{bc}}|^{2}+|\brasket{z_{ac}}{z_{ab}}|^{2}|\brasket{z_{ba}}{z_{bc}}|^{2}}{2 |\braket{z_{ac}}{z_{ab}}\brasket{z_{ac}}{z_{ab}}\braket{z_{ba}}{z_{bc}}\brasket{z_{ba}}{z_{bc}}|} = 
\frac{\cos\phi_{ab}^{c}+\cos\phi_{cb}^{a}\cos\phi_{ac}^{b}}{\sin\phi_{cb}^{a}\sin\phi_{ac}^{b}}\ ,
\end{equation}
where in the last equality we explicited the spinorial scalar products in terms of the 3D dihedral angles $\phi^a_{bc}$ between framed planes $\ket{z_{ab}}$ and $\ket{z_{ac}}$ (see Appendix~\ref{app:aspinor}).

The readers familiar with 4D geometry will recognize in \eqref{eq:spherical_cosine_law} the definition of the angle between the hyperplanes $a$ and $b$ when embedded in 4D, reconstructed using the spherical cosine laws. Without the hat, we will denote the geometric angle with $\theta_{ab}^c$.\footnote{Please note the \emph{plus} sign in the numerator of \eqref{eq:spherical_cosine_law}. At first glance, it seems to disagree with the analog formula reported in some literature \cite{Dittrich:2008va}. The difference resides in the conventions that we use where \emph{all} the dihedral angles (4D, 3D, 2D) are external. The factor $\pi$ of difference compensates for the different sign. The inverse spherical cosine laws have the opposite sign.} The spherical cosine laws have the same form for Euclidean and Lorentzian signatures of the embedding space. If $|\cos\hat\theta_{ab}^{c}|<1$ it is Euclidean and if $|\cos\hat\theta_{ab}^{c}|>1$ it is Lorentzian. In the first case, the angle is real and $\hat\theta_{ab}^{c}= \theta_{ab}^{c}$. In the last case, both hyperplanes are embedded as spacelike hyperplanes. The angle is imaginary and $i\theta_{ab}^{c}=\hat\theta_{ab}^{c} - \chi_{ab} \pi$  where $\chi_{ab}=0$ ($\cos\hat\theta_{ab}^{c}>1$) for \emph{co-chronal} hyperplanes and $\chi_{ab}=1$ ($\cos\hat\theta_{ab}^{c}<1$) for \emph{anti-chronal} hyperplanes. 

The generalized dihedral angle \eqref{eq:spherical_cosine_law} depends only on absolute values of spinors scalar products. On the contrary, the \emph{twist angle} $\xi_{ab}^{c}$ depends explicitely on the phases of the spinors
\begin{equation}
\label{eq:twistangle}
\xi_{ab}^{c}= 
\arg\left(\frac{
\brasket{z_{ac}}{z_{ab}}\braket{z_{ab}}{z_{ac}}
}{
\braket{z_{bc}}{z_{ba}}\sbraket{z_{ba}}{z_{bc}}
}\right) \ .
\end{equation}
It is the edge-dependent and gauge invariant twist angle studied extensively in the \emph{twisted geometry} literature \cite{Dittrich:2008va,Freidel:2010aq,Dittrich:2010ey,Anza:2014tea,Langvik:2016hxn}. It measures the twist between the frames of the planes $ab$ and $ba$ (source and target of the wedge $ab$) using $ac$ and $bc$ as a reference when parallel transported in the same frame with the Ashtekar holonomy. You can follow these steps if you want to visualize the twist angle. Consider the framed plane $\ket{z_{ab}}$ and its intersection with the framed plane $\ket{z_{ac}}$. Do the same with $\ket{z_{ba}}$ and $\ket{z_{bc}}$. Rotate all the framed planes in the edge $b$ such that the framed plane $\ket{z_{ba}}$ coincide with $\ket{z_{ab}}$ and the frames are aligned. The twist angle measures the angle between the intersections of the framed planes with $\ket{z_{ac}}$ and $\ket{z_{bc}}$ respectively.  

\medskip

The local flatness equations equations \eqref{eq:cosine_equations} and \eqref{eq:sine_equations} constrains both the spinors and the complex angles of the wedge holonomies. First, we obtain the complex angles as a function of the spinors by inverting the trigonometric functions. Second, we realize that we have multiple expressions for the same complex angles (one for each 3-cycle the wedge holonomy belongs to). These multiple expressions are consistent if the spinors satisfy some extra conditions. 

The set of equations has been studied in the context of the asymptotic analysis of the EPRL vertex amplitude. We summarize the result here and refer to \cite{Dona:2020yao} for a detailed analysis. We find two distinct classes of solutions.

The first one is characterized by $|\cos\hat\theta_{ab}^{c}|<1$ for all cycles and the angles $\hat\theta_{ab}^{c}=\theta_{ab}^{c}$ are real. Inverting the trigonometric functions in \eqref{eq:cosine_equations} we find
\begin{equation}
\omega_{ab} = i\epsilon\theta_{ab}^c - i\xi_{ab}^c \ , 
\end{equation}
were $\epsilon = \pm 1$ is a sign and has to be the same for all the wedges $ab$ to satisfy \eqref{eq:sine_equations}. There are two ways to have the complex angle cycle independent. If the framed planes satisfy the \emph{orientation conditions}\footnote{All the source and the target framed planes can be oriented in such a way their normals are pairwise opposite. The orientation can happen by rotating all the framed planes at the same edge with the same $SU(2)$ transformation. See \cite{Dona:2017dvf} for complete geometric characterization.} we can prove that $\theta_{ab}^{c} = \xi_{ab}^{c}$ for any $c$.
Therefore, for $\epsilon =1$ the complex angle is trivially cycle independent
\begin{equation}
\omega_{ab} = 0 \, \  \forall ab \ 
\end{equation}
is cycle independent. If $\epsilon =1$ the combination $-\theta_{ab}^c - \xi_{ab}^c$ is not in general independent from the choice of $c$. However, if the spinors satisfy \emph{angle matching} conditions\footnote{Consider the 2D angle $\alpha_{ab}^{cd}$ on the plane $\ket{z_{ab}}$ indentified by its intersection with the planes $\ket{z_{ac}}$ and $\ket{z_{ad}}$. Similarly, conder the corresponding 2D angle $\alpha_{ba}^{cd}$ identified by $\ket{z_{ba}}$, $\ket{z_{bc}}$, and $\ket{z_{bd}}$. The angle matching condition is $\alpha_{ab}^{cd} = \alpha_{ba}^{cd}$ for all possible $a$, $b$, $c$,$d$.} the angles $\theta_{ab}^{c}$ and $\xi_{ab}^{c}$ are individually independent form $c$ (see \cite{Dona:2020yao,Dona:2018nev,Dittrich:2008va,Dittrich:2010ey} for an explicit proof). In this case we drop the cycle dependent index $c$ and the complex angles are purely imaginary and assume two possible values 
\begin{equation}
\omega_{ab} = i\epsilon\theta_{ab} - i\xi_{ab} \ .
\end{equation}
In this case, we can imagine a 4D Euclidean embedding the the edge hyperplanes and the angle $\theta_{ab}$ is the 4D dihedral angle between them in this embedding. The readers familiar with the spin foam literature are probably asking themselves where are vector geometries and the euclidean 4-simplices. We will answer this question at the end of the section. In both cases the complex angles are vanishing or purely imaginary and the wedge holonomies are $SU(2)$ elements. Some parts of the literature refer to this as degenerate geometries. However, we prefer to call this class of solution the \emph{topological sector} to highlight the $SU(2)$ nature of the holonomy. In fact, if we restrict the EPRL spin foam partition function to $SU(2)$ holonomies we obtain the partition function of the topological $SU(2)$ BF theory.

\medskip

The second class of solutions is characterized by $|\cos\hat\theta_{ab}^{c}|>1$ for all cycles. The angles $\hat\theta_{ab}^{c}=i\theta_{ab}^{c}+ \chi_{ab}\pi$ are complex. Inverting the trigonometric functions in \eqref{eq:cosine_equations} we find
\begin{equation}
\label{complexangle}
\omega_{ab} = \epsilon\theta_{ab}^{c} + i \chi_{ab}^{c} \pi - i\xi_{ab}^{c}  \ . 
\end{equation}
The sign $\epsilon = \pm 1$ is the same for all faces $ab$ to satisfy \eqref{eq:sine_equations}. Since it is irrelevant, we dropped the sign from the term containing $\chi_{ab}^{c}$. Because the angle is complex, the only way to have the complex angle \eqref{complexangle} independent from $c$ is to have each term independent from $c$. Once again, if the spinors satisfy \emph{angle matching} conditions the angles $\theta_{ab}^{c}$, $\chi_{ab}^{c}$ and $\xi_{ab}^{c}$ are independent from $c$ \cite{Dona:2020yao,Dona:2018nev,Dittrich:2010ey}. In this case, we drop the cycle-dependent index $c$, and the complex angles assume two possible values 
\begin{equation}
\omega_{ab} = \epsilon\theta_{ab} + i \chi_{ab} \pi - i\xi_{ab} \ .
\end{equation}
We embed the edge hyperplanes in a 4D Lorentzian space as spacelike hyperplanes. We call this sector of solutions the \emph{lorentzian sector}. The real part of the complex angle $\epsilon\theta_{ab}$ is the 4D dihedral angle between the hyperplanes $a$, $b$ embedded in flat Lorentzian 4D space. The imaginary part has a causal term $\chi_{ab} \pi$ that tracks if the hyperplanes are co-chronal or anti-chronal, and a term $\xi_{ab}$ measuring the twist between the framed planes ${ab}$ and ${ba}$. The wedge holonomy has a boost and rotation part and takes the form
\begin{equation}
\label{eq:solutions}
g_{ab} = e^{\frac{\epsilon\theta_{ab} + i \chi_{ab} \pi - i\xi_{ab}}{2}} \sket{z_{ba}}\bra{z_{ab}} - e^{-\frac{\epsilon\theta_{ab} + i \chi_{ab} \pi - i\xi_{ab}}{2}} \ket{z_{ba}}\sbra{z_{ab}} \ .
\end{equation}
The solution with different signs $\epsilon=\pm 1$ are related by the transformation $g_{ab} \stackrel{\epsilon \to -\epsilon}{\longrightarrow} (g_{ab}^\dagger)^{-1}$. This duality is related by the symmetry of the local flatness equations \eqref{eq:3cycleflatness}. We obtain a set of equivalent equations if we transform all the holonomies into their inverse conjugate transpose.

\medskip

Some clarifications are in order. We were cautious in naming the geometries emerging from the local flatness conditions \eqref{eq:3cycleflatness}. In the spin foam literature the geometries in the \emph{topological sector} called \emph{vector geometries}, \emph{Euclidean 4-simplices} and the geometries in the  \emph{Lorentzian sector} are \emph{Lorentzian 4-simplices}. If we impose only \emph{local flatness} we are missing a crucial ingredient to do the same. We do not know if the framed planes at the edges form a tetrahedron. This can be done by adding a closure condition to each edge. 

First, we associate each edge with an area. In the EPRL spin foam model, this area is quantized and given by a spin $j_{ab}=j_{ba}$. We will use the same name for coherence. For each edge $a$ we impose a closure constraint on the spinors at that edge
\begin{equation}
\label{eq:closure}
\sum_{b\neq a} j_{ab} \bra{z_{ab}} \vec{\sigma} \ket{z_{ab}} = 0 \ .
\end{equation}  
With this additional condition the framed planes at each edge $a$ close forming a framed tetrahedron \cite{Livine:2013tsa} with areas given by $j_{ab}$. If the tetrahedra exist, then the geometries in the topological sector satisfying the orientation conditions are vector geometries. The angle-matching condition is equivalent to proper shape-matching conditions since the areas of corresponding triangles match by construction. The geometries in the topological sector satisfying the shape-matching conditions form a Euclidean 4-simplex. The geometries in the Lorentzian sector satisfying the shape-matching conditions form a Lorentzian 4-simplex.
The closure conditions \eqref{eq:closure} are defined up to a global rescaling of all the areas $j_{ab} \to \lambda j_{ab}$. Therefore, the geometries we reconstruct are defined up to a global scale.

%%%%%%%%%%%%%%%%%%%%%%%%%%%%%%%%%%%%%%%%%%%%%%%%%%%%%%%%%%%%%%%%%%%%%%%%%%%%%%%%%%%%%%%%%%

\section{The EPRL model and local flatness}
\label{sec:EPRLdefinition}

%%%%%%%%%%%%%%%%%%%%%%%%%%%%%%%%%%%%%%%%%%%%%%%%%%%%%%%%%%%%%%%%%%%%%%%%%%%%%%%%%%%%%%%%%%
The EPRL spin foam transition amplitude on a given simplicial 2-complex $\Delta$ with faces colored by $SU(2)$ spins $j_f$ and edges colored by intertwiners $i_e$ is given by 
\begin{equation}
\label{eq:spinfoamamplitude}
A_{\Delta} = \sum_{j_f, i_e}  \prod_f A_f \prod_e A_e \prod_v A_v \ .
\end{equation}
The face amplitude is given by $A_f = 2 j_f +1$ and the edge amplitude by $A_e=2i_e+1$. They are fixed \cite{Bianchi:2010fj} requiring the correct convolution property of the path integral with a fixed boundary. The simplicial EPRL vertex amplitude $A_v$ in the spinorial basis is formulated associating to each wedge $ab$ a spin $j_{ab}$ and the function
\begin{equation}
\label{eq:gsimpleirrep}
D_{j_{ab}\mathcal{J}\zeta_{ba}j_{ab}\zeta_{ab}}^{(\gamma j_{ab},j_{ab})}(g_{ab}) \ ,
\end{equation}
where $g_{ab}$ is the wedge holonomy, $D_{j_{ab}\mathcal{J}\zeta_{ba}j_{ab}\zeta_{ab}}^{(\gamma j_{ab},j_{ab})}$ is a $\gamma$-simple unitary irreducible representation of $SL(2,\C)$, and $\gamma$ is the Immirzi parameter. The choice of the particular $\gamma$-simple representation is due to the weak quantum implementation of the linear simplicity constraints. Classically, it is responsible for reducing the Lorentzian topological BF theory to general relativity.
In the spinorial basis, the $\gamma$-simple irrep \eqref{eq:gsimpleirrep} can be written in exponential form using a dummy spinor $w_{ab}$ necessary to implement the unitarity of the representation
\begin{equation}
\label{eq:wedgeexp}
D_{j_{ab}\mathcal{J}\zeta_{ba}j_{ab}\zeta_{ab}}^{(\gamma j_{ab},j_{ab})}(g_{ab})=\frac{2j_{ab}+1}{\pi}\int \d w_{ab}
\frac{\sbraket{\zeta_{ba}}{w_{ab}}^{2j_{ab}}\bra{w_{ab}}g_{ab}\ket{\zeta_{ab}}^{2j_{ab}}}{\Vert w_{ab}\Vert ^{2j_{ab}+2i\gamma j_{ab}+2} \Vert g_{ab}^{\dagger}w_{ab}\Vert ^{2j_{ab}-2i\gamma j_{ab}+2}} = \int \d w_{ab} \mu_{ab} e^{S_{ab}} \ ,
\end{equation}
where the function $\mu_{ab} = \frac{2j_{ab}+1}{\pi} \frac{1}{\Vert w_{ab}\Vert^{2} \Vert g_{ab}^{\dagger}w_{ab}\Vert^{2}}$ is usually interpreted as a measure factor. The wedge action is 
\begin{equation}
\label{eq:wedgeaction}
S_{ab} = 2j_{ab} \log \left( \frac{\sbraket{\zeta_{ba}}{w_{ab}}\bra{w_{ab}}g_{ab}\ket{\zeta_{ab}}}{\Vert w_{ab}\Vert ^{1+i\gamma} \Vert g_{ab}^{\dagger}w_{ab}\Vert ^{1-i\gamma}}\right) \ .
\end{equation}
The vertex amplitude $A_v$ is constructed taking the product of \eqref{eq:gsimpleirrep} on all the wedges. Then, assigning a holonomy to each edge $g_a$ such that $g_{ab} = g_{b}^{-1}g_{a}$. And finally, integrating over the edge holonomies, being careful to remove a redundant integration to regularize the amplitude \cite{Engle:2008ev}.
\begin{equation}
\label{eq:EPRLvertex}
A_v = \int \prod_{a} \d g_{a}  \delta(g_1) \prod_{ab}  D_{j_{ab}\zeta_{ba}j_{ab}\zeta_{ab}}^{(\gamma j_{ab},j_{ab})}(g_{b}^{-1}g_{a}) \ .
\end{equation}
The model is usually introduced in this way, and the literature rarely comments on the choice of assigning holonomies to edges. Introducing the edge holonomy $g_a$ related to the wedge holonomy via $g_{ab} = g_{b}^{-1}g_{a}$ implements the \emph{local flatness} condition strongly in the EPRL model. 
Local flatness is not a requirement of the discretization prescription we use in formulating the EPRL model. It is an ingredient we add to the mix because we want to work with flat simplices. To highlight it, we rewrite the vertex amplitude \eqref{eq:EPRLvertex} using all ten wedge holonomies and imposing \emph{local flatness} strongly with a delta function for each fundamental cycle of the 4-simplex. 
\begin{equation}
\label{eq:EPRLvertex_lf}
A_v = \int \left(\prod_{ab} \d g_{ab}   D_{j_{ab}\zeta_{ba}j_{ab}\zeta_{ab}}^{(\gamma j_{ab},j_{ab})}(g_{ab}) \right) \mathcal{C}_{LF}(g_{ab},\cdots,g_{cd}) %\prod_{\mathcal{C} \in \mathcal{F}} \delta(\prod_{w\in \mathcal{C}}g_{w})\ .
\end{equation}
An explicit example of $\mathcal{C}_{LF}$ based on a choice of fundamental cycles is
\begin{equation}
\label{eq:CLF}
\begin{split}
\mathcal{C}_{LF}(g_{ab},\cdots,g_{cd}) =& \delta\left(g_{13}^{-1}g_{23}g_{12}\right)\delta\left(g_{14}^{-1}g_{24}g_{12}\right)\delta\left(g_{15}^{-1}g_{25}g_{12}\right)  \\
									 & \delta\left(g_{14}^{-1}g_{34}g_{13}\right)\delta\left(g_{15}^{-1}g_{35}g_{13}\right)\delta\left(g_{15}^{-1}g_{45}g_{14}\right) \ .
\end{split}
\end{equation}
An explicit integration shows that we can eliminate six wedge holonomies $g_{ab}$ with $a > 1$ from \eqref{eq:EPRLvertex_lf} out of ten. We obtain exactly the formula of the amplitude \eqref{eq:EPRLvertex} if we identify the edge holonomies $g_b$ with $g_{1b}$. Different choices of fundamental cycles in \eqref{eq:CLF} correspond to different regularizations of \eqref{eq:EPRLvertex} where a different redundant edge integration is removed.

%%%%%%%%%%%%%%%%%%%%%%%%%%%%%%%%%%%%%%%%%%%%%%%%%%%%%%%%%%%%%%%%%%%%%%%%%%%%%%%%%%%%%%%%%%

\section{The large spin limit of the EPRL vertex amplitude}
\label{sec:EPRLsaddle}

%%%%%%%%%%%%%%%%%%%%%%%%%%%%%%%%%%%%%%%%%%%%%%%%%%%%%%%%%%%%%%%%%%%%%%%%%%%%%%%%%%%%%%%%%%
In the large spin regime, we consider the spins associated with the faces EPRL vertex amplitude homogenously large
\begin{equation}
\label{eq:largespins}
j_{ab} \to \lambda j_{ab} \qquad\text{with}\qquad \lambda \gg 1 \ .
\end{equation}
We will refer to $\lambda$ as the scale of the spins. We recast the vertex amplitude in exponential form using \eqref{eq:wedgeexp} with a total action
\begin{equation}
\label{eq:totalaction}
S = \sum_{ab} S_{ab} = \sum_{ab} 2j_{ab} \log \left( \frac{\sbraket{\zeta_{ba}}{w_{ab}}\bra{w_{ab}}g_{ab}\ket{\zeta_{ab}}}{\Vert w_{ab}\Vert ^{1+i\gamma} \Vert g_{ab}^{\dagger}w_{ab}\Vert ^{1-i\gamma}}\right) \ .
\end{equation}
The action \eqref{eq:totalaction} is linear in the spins. Therefore, we can approximate the vertex amplitude using saddle point techniques. The saddle points dominate the integrals of the vertex amplitude and are the points where the gradient of the action vanishes.
Among them, we focus on those maximizing the real part of the action and giving the dominant contributions to the asymptotic behavior of the vertex amplitude. The real part of the wedge actions is negative $\Re S_{ab}\leq 0$. The maximum of the real part of whole action \eqref{eq:totalaction} is obtained when $\Re S_{ab} = 0$ for all wedges.
Strictly speaking, we should consider all the saddle points. Traditionally, the dominant ones are studied since the additional equations simplify the calculations. 
It would be interesting to study what changes if we relax the dominance request, especially in extended triangulations \cite{Han:2011re,Han:2011rf,Asante:2022lnp}. The saddle point equations are
\begin{eqnarray}
\label{eq:groupvariation}
  \frac{\delta S}{\delta g_a} = 0\\
\label{eq:dummyvariation}
  \frac{\delta S}{\delta w_{ab}} = 0\\
\label{eq:dummyvariationcc}
  \frac{\delta S}{\delta \bar{w}_{ab}} = 0\\
\label{eq:ReSequal0}
  \Re S_{ab} = 0 
\end{eqnarray}
We combine the saddle point equations to recast them in a form with a simpler geometrical interpretation. From \eqref{eq:groupvariation} using the other equations to remove the dummy spinors $w_{ab}$, we obtain the closure condition for the boundary spinors
\begin{equation}
\label{eq:closure_bdry}
\sum_{b\neq a} j_{ab} \bra{\zeta_{ab}} \vec{\sigma} \ket{\zeta_{ab}} = 0 \ .
\end{equation}
As we already discussed in \eqref{eq:closure} this condition allows interpreting the boundary framed planes as framed tetrahedra with areas given by the spins $j_{ab}$. 

Combining the other three equations \eqref{eq:ReSequal0}, \eqref{eq:dummyvariation} and \eqref{eq:dummyvariationcc}, we obtain some equations that can be used to determine the value of the dummy spinors at the saddle point (we will not report them since they are not important for our discussion) and the equations that constrain the wedge holonomies' spinors in terms of the boundary ones. We use wedge holonomies to separate the saddle point equations from the local flatness equation $g_{ab}$ instead of the edge holonomies. 
\begin{equation}
\label{eq:alignment}
  g_{ab} \ket{\zeta_{ab}} = \lambda_{ab} \sket{\zeta_{ba}} \ ,  \qquad (g_{ab}^{-1})^\dagger \ket{\zeta_{ab}} = \frac{1}{\lambda_{ab}^*} \sket{\zeta_{ba}} \ ,
\end{equation}
where the proportionality parameter $\lambda_{ab}$ is to be determined. We have one alignment equation for each wedge $ab$ fixing the parameters of the wedge holonomies \eqref{eq:wedgeholo} in terms of the boundary states.
\begin{equation}
\label{eq:onshellholonomy}
g_{ab} = e^{\frac{\omega_{ab}}{2}} \sket{\zeta_{ba}}\bra{\zeta_{ab}} - e^{-\frac{\omega_{ab}}{2}} \ket{\zeta_{ba}}\sbra{\zeta_{ab}} \ ,
\end{equation}
where the porportionality coefficient is related to the complex angle $\lambda_{ab} = e^{\tfrac{\omega_{ab}}{2}}$ and still free.\footnote{The proof is straightforward. We write $g_{ab}$ in terms of the four bilinears $\ket{\zeta_{ba}}\bra{\zeta_{ab}}$, $\sket{\zeta_{ba}}\bra{\zeta_{ab}}$, $\ket{\zeta_{ba}}\sbra{\zeta_{ab}}$, $\sket{\zeta_{ba}}\sbra{\zeta_{ab}}$ that form a basis of $GL(2,\C)$ and we use \eqref{eq:alignment} and $\det g_{ab} =1$ to determine the complex coefficients.} 
%
%\medskip
%
%The alignment equations \eqref{eq:alignment} support the interpretation of the framed plane $\ket{\zeta_{ab}}$ $\gamma$-simple bivectors in 4D Minkowski space. As described in the previous section we can associate to each spinor a $\gamma$-simple bivector $\Pi_{ab} = j_{ab}(1 + i \gamma) \bra{\zeta_{ab}}\vec{\sigma} \ket{\zeta_{ab}}$ and $\Pi_{ba}  = j_{ab}(1 + i \gamma) \bra{\zeta_{ba}}\vec{\sigma} \ket{\zeta_{ba}}$. Using both equations \eqref{eq:alignment} we find that $g_{ab}$ transform one $\gamma$-simple bivector into the other $\Pi_{ba} = g_{ab} \triangleright \Pi_{ab}$.

\medskip

On-shell of the alignment equations \eqref{eq:alignment} the action \eqref{eq:totalaction} assumes a very simple form in terms of the complex angles $\omega_{ab}$
\begin{equation}
S^{(cr)} = \sum_{ab} S_{ab}^{(cr)} = i \lambda \sum_{ab}  j_{ab} \left( \gamma  \Re \omega_{ab} + \Im \omega_{ab} \right) \ .
\end{equation}

Suppose we do not impose the local flatness conditions. In that case, the complex angles in \eqref{eq:onshellholonomy} are undetermined, and the closure conditions \eqref{eq:closure} are the only restrictions on the boundary spinors. To connect with discrete 4D geometries and reproduce the famous result for which Lorentzian Regge action (and geometry) emerge in the semiclassical limit of the EPRL vertex amplitude, we have to include the local flatness conditions that we discussed in great length in section \eqref{sec:localflatness}. They require additional constraints on the spinors and fix the complex angles in terms of the holonomies' spinors. For example, if we restrict to the Lorentzian sector, local flatness requires the framed tetrahedra described by the boundary spinors to satisfy the shape-matching constraints. We can reconstruct a Lorentzian 4-simplex with a spacelike boundary embedding the framed tetrahedra in 4D. The real part of the complex angles equals the dihedral angles. The imaginary part of the complex angles relates to the twist angle between the two framed tetrahedra. The action at the saddle point in this case becomes
\begin{equation}
S^{(cr)} = i \lambda \epsilon \sum_{ab}  \gamma j_{ab}   \theta_{ab} - i \lambda \sum_{ab}  j_{ab} \xi_{ab} +  i \lambda \epsilon \pi  \sum_{ab}  j_{ab} \chi_{ab} = 
i \lambda \epsilon S_R - i \lambda T +  i \lambda \epsilon \pi \Upsilon \ .
\end{equation}
The first term is the well known Regge action of the 4-simplex $S_R = \sum_{ab} \gamma j_{ab} \theta_{ab}$ with areas $\gamma j_{ab}$ and Lorentzian dihedral angles $\theta_{ab}$. The sign $\epsilon$, which labels the two solutions of the local flatness equations, represents the two possible orientations of the 4-simplex with given boundary spacelike tetrahedra. The second term $T=\sum_{ab}  j_{ab} \xi_{ab}$ is a twist term describing how the holonomies transform the frames of the boundary triangles. The last term $\Upsilon= \sum_{ab}  j_{ab} \chi_{ab}$ is related to the ``causal'' structure of the 4-simplex and encodes the co- or anti-chronality of the boundary tetrahedra in the 4-simplex. 

%%%%%%%%%%%%%%%%%%%%%%%%%%%%%%%%%%%%%%%%%%%%%%%%%%%%%%%%%%%%%%%%%%%%%%%%%%%%%%%%%%%%%%%%%%

\section{Extended EPRL spin foam amplitudes}
\label{sec:manyvertices}

\begin{figure}[H]
\centering
\includegraphics[scale=0.7]{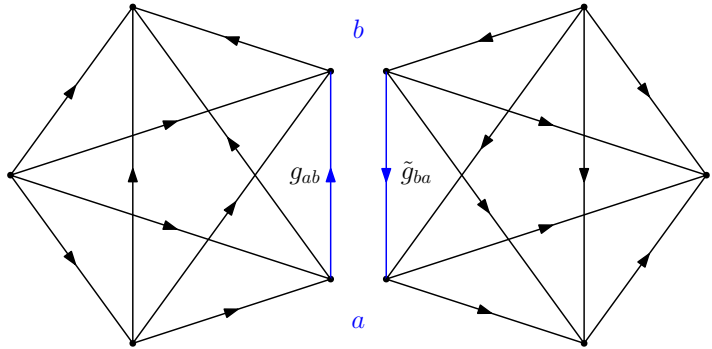}
\caption{\label{fig:two-vertices} Two spin foam vertices glued together. We denoted with $a$ and $b$ the bulk edges (highlighted in blue). The bulk face (also highlighted in blue) connects the two vertices and comprises the two wedges (one for each vertex) that include the bulk edges. We indicate the two bulk holonomies with $g_{ab}$ and $\tilde g_{ba}$.}
\end{figure}

%%%%%%%%%%%%%%%%%%%%%%%%%%%%%%%%%%%%%%%%%%%%%%%%%%%%%%%%%%%%%%%%%%%%%%%%%%%%%%%%%%%%%%%%%%
The next step is considering spin foam amplitudes with more than a vertex. Before discussing the amplitude of the EPRL model specifically, we will make some general observations. For simplicity of notation, let us consider the most straightforward case of two vertices glued together on two edges. The 2-complex associated with this amplitude has two bulk edges and a bulk face (see Figure~\ref{fig:two-vertices} for reference). We decorate with tides the variables associated with the second vertex and look at the holonomies related to the bulk face. 
Using the parametrization \eqref{eq:wedgeholo} the two bulk holonomies are
\begin{equation}
\label{eq:twoholo}
g_{ab} = e^{\frac{\omega_{ab}}{2}} \sket{z_{ba}}\bra{z_{ab}} - e^{-\frac{\omega_{ab}}{2}} \ket{z_{ba}}\sbra{z_{ab}} \ , \qquad \text{and} \qquad
\tilde g_{ba} = e^{\frac{\tilde \omega_{ba}}{2}} \sket{\tilde z_{ab}}\bra{\tilde z_{ba}} - e^{-\frac{\tilde \omega_{ba}}{2}} \ket{\tilde z_{ab}}\sbra{\tilde z_{ba}} \ .
\end{equation}
In general, the two holonomies in their \eqref{eq:wedgeholo} form do not glue nicely. The framed plane $\ket{z_{ba}}$ of $g_{ab}$ is independent from the framed plane $\ket{\tilde z_{ba}}$ of $\tilde g_{ba}$. The details of how the two holonomies glue is given by the details of the spin foam theory.

\medskip 

We focus on the contribution to the EPRL spin foam amplitude $A_\Delta$ \eqref{eq:spinfoamamplitude} deriving from the bulk face. Each vertex contributes with a $\gamma$-simple unitary irreducible representation \eqref{eq:gsimpleirrep}
\begin{equation}
\label{eq:facecontrib-noexp}
D_{j_{ab}\mathcal{J}\zeta_{ba}j_{ab}\zeta_{ab}}^{(\gamma j_{ab},j_{ab})}(g_{ab}) \ , \qquad \text{and} \qquad
D_{j_{ab}\mathcal{J}\tilde\zeta_{ab}j_{ab}\tilde\zeta_{ba}}^{(\gamma j_{ab},j_{ab})}(\tilde g_{ba}) \ .
\end{equation}
We glue the two vertices together with the factors $\brasket{\zeta_{ab}}{\tilde\zeta_{ab}}^{2j_{ab}}$ and $\brasket{\zeta_{ba}}{\tilde\zeta_{ba}}^{2j_{ab}}$ due to the fact that the spinorial basis is overcomplete. We integrate over the four spinors $\zeta$ with the appropriate measure
\footnote{The integration measure is $\d\zeta = 1/pi \d^4 \zeta$.}. If we put \eqref{eq:facecontrib-noexp} in the exponential form we find that the face contributes to the amplitude with an action 
\begin{equation}
\label{eq:facecontribution}
\begin{split}
S_f =& 2j_{ab} \log \left( \frac{\sbraket{\zeta_{ba}}{w_{ab}}\bra{w_{ab}}g_{ab}\ket{\zeta_{ab}}}{\Vert w_{ab}\Vert ^{1+i\gamma} \Vert g_{ab}^{\dagger}w_{ab}\Vert ^{1-i\gamma}}\right) +  2j_{ab} \log \brasket{\zeta_{ab}}{\tilde\zeta_{ab}} + \\
& 2j_{ab} \log \left( \frac{\sbraket{\tilde \zeta_{ab}}{\tilde w_{ab}}\bra{\tilde w_{ab}}\tilde g_{ba}\ket{\tilde \zeta_{ba}}}{\Vert \tilde w_{ab}\Vert ^{1+i\gamma} \Vert \tilde g_{ab}^{\dagger}\tilde w_{ab}\Vert ^{1-i\gamma}}\right) + 2j_{ab} \log \brasket{\tilde \zeta_{ba}}{\zeta_{ba}} \ .
\end{split}
\end{equation}
All the spinors in \eqref{eq:facecontribution} are bulk spinors used to glue the vertices or dummy spinors used i nthe $SL(2,\C)$ representations. We are integrating over all of them. %In the 2-complex we are considering all the other faces are boundary faces. Their contribution is similar to \eqref{eq:facecontribution} and \eqref{eq:wedgeaction} and to not overload the reader with details we will not report explicit formulas. \comment{Maybe appendix?}

%. We can distinguish two kind of them: faces that include a bulk edge and faces that do not. In the first case the contribution of the face to the amplitude is with an action  
%\begin{equation}
%\label{eq:bdryfacecontribution}
%\begin{split}
%S_f =& 2j_{ac} \log \left( \frac{\sbraket{\zeta_{ca}}{z_{ac}}\bra{z_{ac}}g_{ac}\ket{\zeta_{ac}}}{\Vert z_{ac}\Vert ^{1+i\gamma} \Vert g_{ac}^{\dagger}z_{ac}\Vert ^{1-i\gamma}}\right) +  2j_{ac} \log \brasket{\zeta_{ac}}{\tilde\zeta_{ac}} +  2j_{ac} \log \left( \frac{\sbraket{\tilde \zeta_{ac}}{\tilde z_{ac}}\bra{\tilde z_{ac}}\tilde g_{ca}\ket{\tilde \zeta_{ca}}}{\Vert \tilde z_{ac}\Vert ^{1+i\gamma} \Vert \tilde g_{ac}^{\dagger}\tilde z_{ac}\Vert ^{1-i\gamma}}\right) \ .
%\end{split}
%\end{equation}
%The spinors $\ket{\zeta_{ac}}$ and $\ket{\tilde\zeta_{ac}}$ are integrated over while the other two $\ket{\zeta_{ca}}$ and $\ket{\tilde\zeta_{ca}}$ are boudnary data. In the second case the contribution is identical to the one of a single vertex \eqref{eq:wedgeaction} involving only two boundary edges.
The whole EPRL amplitude \eqref{eq:spinfoamamplitude} is defined with a sum over spins associated with bulk faces. Before considering the effect of the sum, we analyze the amplitude for a fixed value of the bulk face spin. 

The action \eqref{eq:facecontribution} is still linear in the spins $j_{ab}$ and in the large spins regime \eqref{eq:largespins} we can approximate the spin foam amplitude using saddle point techniques. The saddle point is made on all the variables. Now they also include bulk spinors. Again we look for the dominant saddle points where the gradient of the action vanishes, and the real part of the action is maximal. Since $\Re S \leq 0$ and both the contribution coming from the vertices and the gluing are negative and at most zero, maximizing $\Re S$ is equivalent to asking that the real part of each of its constituents (each logarithm) vanish independently. 

In addition to the saddle point equations we discussed in Section~\ref{sec:EPRLsaddle} for each vertex, the extra integrals over the bulk spinors lead to some additional equations

\begin{eqnarray}
\label{eq:exbdryvariation}
\frac{\delta S}{\delta \zeta_{ab}} =0 \ , \\
\label{eq:exbdryvariationcc}
\frac{\delta S}{\delta \bar\zeta_{ab}} =0 \ , \\
\label{eq:scal}
  \Re \log \brasket{\zeta_{ab}}{\tilde\zeta_{ab}} =  0 \ .
\end{eqnarray}

It has been shown with an explicit calculation \cite{Han:2011re,Han:2011rf} that on-shell of the other critical point equations \eqref{eq:exbdryvariation} and \eqref{eq:exbdryvariationcc} are automatically satisfied. The solution of the equations \eqref{eq:scal} are simple and requires the bulk edge spinors to coincide up to a phase
\begin{equation}
\label{eq:gluing}
\ket{\zeta_{ab}} = e^{-i\alpha_{ab}}\sket{\tilde\zeta_{ab}} \ , \qquad \text{and} \qquad \ket{\zeta_{ba}} = e^{-i\alpha_{ba}}\sket{\tilde\zeta_{ba}} \ .
\end{equation}

The saddle point analysis is identical to the single vertex case with the extra wedge gluing condition on bulk edge spinors. The action of the closed face \eqref{eq:facecontribution} at the critical points is 
\begin{equation}
S_f^{(cr)} =  i \lambda j_{ab} \left( \gamma  \Re \omega_{ab} + \Im \omega_{ab} \right) +  i \lambda j_{ab} \left( \gamma  \Re \tilde\omega_{ab} + \Im \tilde\omega_{ab} \right)  + i 2 \lambda j_{ab} (\alpha_{ab} + \alpha_{ba})\ .
\end{equation}

To make contact with geometry, we go on-shell of each vertex's local flatness equations \eqref{eq:3cycleflatness}. If we restrict for brevity to a global Lorentzian sector\footnote{While critical point equations (the spherical sine laws, to be precise) force all the holonomies in the same vertex to be in the same sector, there are no equations that require holonomies in different vertices to be in the same sector.}, the action of the closed face \eqref{eq:facecontribution} reduces to 
\begin{equation}
S_f^{(cr)} =  
i \lambda \gamma j_{ab}  \left( \epsilon \theta_{ab} +  \tilde\epsilon  \tilde\theta_{ba}\right) 
- i \lambda j_{ab} \left(  \xi_{ab} + \tilde \xi_{ba}\right)  
+ i 2 \lambda j_{ab} (\alpha_{ab} + \alpha_{ba})
+ i \lambda \pi j_{ab} \left( \epsilon  \chi_{ab}  + \tilde \epsilon \chi_{ba}\right) \ .
\end{equation}
Recall that we decorate with a tilde the variables of the second vertex. The first term is the algebraic sum of the Lorentzian dihedral angle dual to the triangle $(ab)$. It is related to the deficit angle associated with the triangle, and the relative sign $\epsilon / \tilde \epsilon$ is associated with the relative orientation of the two 4-simplices. The relative orientation of the two vertices is not fixed. If this is a desired feature of the EPRL model or if we want to remove it is not clear; however, the same freedom is present in the Ponzano-Regge model \cite{Regge:1968,Dona:2019jab} where it is necessary to have a topological invariant model.  

The last term is a sign for the amplitude, keeping track of the local causal structures of the two 4-simplices. The twist term combines with the overlapping term in an interesting way. Observe that, up to irrelevant factors of $2\pi$, and using the conditions \eqref{eq:gluing} for the closed face and the auxiliary face used to define the twist angles, we can express the sum of the twist angles as 
\begin{equation}
\label{eq:twistsimplified}
\xi_{ab}+\tilde \xi_{ab}=
\arg\left(\frac{
\brasket{\zeta_{ac}}{\zeta_{ab}}\braket{\zeta_{ab}}{\zeta_{ac}}
}{
\braket{\zeta_{bc}}{\zeta_{ba}}\sbraket{\zeta_{ba}}{\zeta_{bc}}
}\right) + 
\arg\left(\frac{
\braket{\zeta_{bc}}{\zeta_{ba}}\sbraket{\zeta_{ba}}{\zeta_{bc}}
}{
\brasket{\zeta_{ac}}{\zeta_{ab}}\braket{\zeta_{ab}}{\zeta_{ac}}
}\right) + 2(\alpha_{ab}+\alpha_{ba}) = 2(\alpha_{ab}+\alpha_{ba}) \ , 
\end{equation}
where the two $\arg$ functions cancel since their arguments are reciprocal. The twist contribution of the action summed with the overlap on-shell vanish. The action of the closed face \eqref{eq:facecontribution} restricted to a global Lorentzian sector reduces to a term proportional to the deficit angle and a ``sign'' term related to the Lorentzian structure of the two 4-simplices
\begin{equation}
S_f^{(cr)} =  
i \lambda \gamma j_{ab}   \left( \epsilon \theta_{ab} +  \tilde\epsilon  \tilde\theta_{ba}\right) 
+ i \lambda \pi j_{ab} \left( \epsilon  \chi_{ab}  + \tilde \epsilon \chi_{ba}\right) \ .
\end{equation}

It is interesting to also look at the form of the face holonomy on-shell of the critical point equations. Using the gluing equations \eqref{eq:gluing} we have \begin{equation}
\label{eq:facehololargespins}
g_f = \tilde g_{ba} g_{ab} = e^{\frac{\omega_{ab}+\tilde \omega_{ba}}{2} + i (\alpha_{ab} + \alpha_{ba})} \ket{\zeta_{ab}}\bra{\zeta_{ab}} + e^{-\frac{\omega_{ab}+\tilde \omega_{ab}}{2}- i (\alpha_{ab} + \alpha_{ba})} \sket{\zeta_{ab}}\sbra{\zeta_{ab}} \ .
\end{equation}
The holonomy $g_f$ maps the framed plane $\ket{z_{ab}}$ to itself with a complex angle $\omega_f = \omega_{ab}+\tilde \omega_{ab} + i 2 (\alpha_{ab} + \alpha_{ba})$. The spinor $\ket{\zeta_{ab}}$ is constrained only by closure equations. In particular, $g_f$ is a 4-screw, a composition of a boost of angle $\Re\omega_f$ and a rotation of angle $\Im\omega_f$ along the same axis $\vec{n}_{ab}$.

If we include the local flatness equations \eqref{eq:3cycleflatness} and restrict ourselves to the global Lorentzian sector, we can furthermore simplify the expression for the complex angle $\omega_f$
\begin{equation}
\omega_f =  
\gamma \left( \epsilon \theta_{ab} +  \tilde\epsilon  \tilde\theta_{ba}\right) + i \pi \left( \epsilon  \chi_{ab}  + \tilde \epsilon \chi_{ba}\right)
 \ ,
\end{equation}
where we used \eqref{eq:twistsimplified} to cancel the twist angles with the gluing phases, in this sector, the face holonomy has the form of a Regge holonomy. It is a pure boost in the plane orthogonal to the triangle of an angle equal to the sum of the 4D dihedral angles dual to the triangles of the (two) 4-simplices that share it.

%%%%%%%%%%%%%%%%%%%%%%%%%%%%%%%%%%%%%%%%%%%%%%%%%%%%%%%%%%%%%%%%%%%%%%%%%%%%%%%%%%%%%%%%%%

\section{The sum over the spins}
\label{sec:sumspins}

%%%%%%%%%%%%%%%%%%%%%%%%%%%%%%%%%%%%%%%%%%%%%%%%%%%%%%%%%%%%%%%%%%%%%%%%%%%%%%%%%%%%%%%%%%
In the previous section, we studied the large spins approximation of the spin foam amplitude with fixed bulk spins. The full spin foam amplitude \eqref{eq:spinfoamamplitude} is obtained by weighting the contributions with a face amplitude $A_f = 2j_f+1$ and summing over the spins associated with bulk faces. To keep the notation simple, we will continue using the example introduced in the previous section. The generalization to arbitrary bulk faces immediately follows with the same reasoning and steps. Summing over its spin explicitly, the bulk face $(ab)$ contributes to the spin foam amplitude with
\begin{equation}
\label{eq:WEPRL}
\begin{split}
F_{EPRL}(g_{ab},\tilde{g}_{ba})=\sum_{j_{ab}} (2j_{ab}+1)& \int (2j_{ab}+1) \d \zeta_{ab} \int (2j_{ab}+1) \d \zeta_{ba} \int (2j_{ab}+1) \d \tilde\zeta_{ab} \int (2j_{ab}+1) \d \tilde\zeta_{ba}  \\
&
 D_{j_{ab}\mathcal{J}\zeta_{ba}j_{ab}\zeta_{ab}}^{(\gamma j_{ab},j_{ab})}(g_{ab})
\brasket{\zeta_{ab}}{\tilde\zeta_{ab}}^{2 j_{ab}}
D_{j_{ab}\mathcal{J}\tilde\zeta_{ab}j_{ab}\tilde\zeta_{ba}}^{(\gamma j_{ab},j_{ab})}(\tilde g_{ba})
\brasket{\tilde\zeta_{ba}}{\zeta_{ba}}^{2 j_{ab}} \ , 
\end{split}
\end{equation}
where $\d \zeta$ is the measure over the spinor space. Notice that the integrals over the spinors could be performed exactly. We opt not to do it to use the same notation of the previous sections. 

\medskip

The partition function of the EPRL spin foam model can be rewritten as the product of functions like \eqref{eq:WEPRL} associated with each face, a local flatness constraint for each vertex, and the integration over all the wedge holonomies. The equivalent of \eqref{eq:WEPRL} for the BF topological spin foam theory is a group delta function of the product of the wedge holonomies associated with the face. In that case, the delta functions impose the global flatness of the model on the 2-complex, forcing any parallel transport to be trivial. A similar result for the EPRL model can be obtained by studying the wavefront set of the partition function. Studying it is a very involved process and has been proposed in the context of the Euclidean EPRL model in \cite{Hellmann:2013gva}. The analysis in \cite{Hellmann:2013gva} focuses more on mathematical rigor than its interpretation and uses a parametrization with numerous auxiliary variables. For these reasons, it remains an amazing paper not fully digested by the community. The takeaway message of their work we need for this section is that the singular support of the integrand dominates the integrals in the EPRL partition functions. The singular support of a distributions is essentially given by the group elements for which the distribution is singular (e.g. the singular support of $\delta(g_f)$ is given by $\mathrm{Sing}(\delta) = \lbrace \mathds{1}\rbrace$). 

\medskip

The calculation is lengthy but simple. We report it with all the details in Appendix~\ref{app:singularsupport}.  The singular support of the function \eqref{eq:WEPRL} is given by 
\begin{equation}
\label{eq:singsuppW}
\mathrm{Sing} (F_{EPRL}(g_{ab},\tilde{g}_{ba})) = \left\lbrace g_{ab},\tilde{g}_{ba}\ \vert \ \tilde{g}_{ba} g_{ab} = 
e^{\frac{\omega_{f}}{2}} \ket{\zeta}\bra{\zeta} + e^{-\frac{\omega_{f}}{2}} \sket{\zeta}\sbra{\zeta} \ \text{with} \ \gamma \Re \omega_f + \Im \omega_f = 0 \ \mathrm{mod} \ 4\pi \right\rbrace \ .
\end{equation}

The singular support of \eqref{eq:WEPRL} contains all the wedge holonomies such that their product (that forms the face holonomy representing the parallel transport around a bulk face) is a 4-screw. Its direction is arbitrary, and it has a rapidity and rotation angle proportional (the proportionality factor is $-\gamma$) up to $4\pi$ factors. Following the arguments of \cite{Hellmann:2013gva}, we expect that such holonomies dominate the semiclassical limit. The result of the analog calculation for a face with an arbitrary number of vertices is the same. 

The ``naive'' flatness problem \cite{Conrady:2008mk,Bonzom:2009hw,Han:2013ina,Hellmann:2013gva,Engle:2020ffj}  claims that, at fixed triangulation, the amplitude is dominated in the large (boundary) spin limit by flat geometries.
For a path integral formulation of a quantum theory, the amplitude is exponentially suppressed if and only if the boundary data is inconsistent with the classical equation of motions. Therefore the flatness problem was interpreted as an indication that the EPRL model apparently cannot recover non-flat solutions of Einstein's equations. The problem disappears if the triangulation refinement is considered at the same time as the large spin limit.   

However, we want to show that the same tension arises when we combine \eqref{eq:singsuppW} with the large spin limit form of the face holonomy derived in the previous section \eqref{eq:facehololargespins} and local flatness conditions. In that case, the imaginary part of the complex angle of the face holonomy is the vanishing twist term. The real part of the complex angle is the sum of the dihedral angles associated with the bulk face. In this case, the singular support condition gives the (in)famous flatness equation $\gamma \sum_i \theta_i = 0 \ \mathrm{mod} \ 4\pi $.

The same conclusion can also be derived using saddle point techniques, and Poisson resummation \cite{Engle:2020ffj}. We preferred to follow the singular support argument to demystify the result of \cite{Hellmann:2013gva} because it focuses more on the holonomies than the geometries.

%%%%%%%%%%%%%%%%%%%%%%%%%%%%%%%%%%%%%%%%%%%%%%%%%%%%%%%%%%%%%%%%%%%%%%%%%%%%%%%%%%%%%%%%%%

\section{Discussion}
\label{sec:conclusion}

%%%%%%%%%%%%%%%%%%%%%%%%%%%%%%%%%%%%%%%%%%%%%%%%%%%%%%%%%%%%%%%%%%%%%%%%%%%%%%%%%%%%%%%%%%
We explored the implication of local flatness in Lorentzian spin foam models. We find that geometry emerges naturally from it, independently from the details of the models. 

\medskip

We associate a set of $SL(2,\C)$ holonomies with the wedges of a spin foam vertex. We require the parallel transport within a simplicial spin foam vertex to be trivial and find constraints on the geometry that parametrize the holonomies. We parametrize each holonomy with two framed planes (one at the source and one at the target) and a complex angle representing the boost and twist between the planes. This interpretation is analog to the twisted geometry picture of loop quantum gravity \cite{Freidel:2010aq,Dittrich:2010ey,Speziale:2012nu,Anza:2014tea,Langvik:2016hxn}. The local flatness equations impose conditions on the angles of the geometry that we divide into two classes. The first class of solutions contains $SU(2)$ holonomies. Moreover, the geometry has to satisfy at least the orientation conditions, and the complex angles vanish. If the geometry satisfies angle-matching conditions, we can embed the 3D hyperplanes associated with the spin foam edges in Euclidean 4D space. The complex angles are related to the 4D dihedral angles and the twist between corresponding frames.
The second class of solutions contains holonomies with a non-trivial boost part. The geometry satisfies angle-matching conditions, and we can embed the 3D hyperplanes associated with the spin foam edges as space-like hyperplanes in a Lorentzian 4D space. The real part of the complex angles relates to the 4D dihedral angle between them, and the imaginary part to the twist between corresponding frames.

\medskip

If we require closure conditions of the framed planes associated with the same edge, we can interpret the framed planes as framed tetrahedra. If we supplement the local flatness condition with closure conditions, geometry in the topological sector reduces to a vector geometry or a Euclidean 4-simplex. Geometry in the Lorentzian sector reduces to a Lorentzian 4-simplex with a space-like boundary. Closure conditions result from the theory's edge $SU(2)$ invariance. 
Assume we find a mechanism to select the Lorentzian sector of locally flat holonomies. Integrating over locally flat holonomies satisfying edge closure constraints restricted to the Lorentzian sector is equivalent to summing over all Lorentzian 4-simplices.

\medskip

The correspondence between holonomies and Lorentzian geometry is model-independent and very general. We do not have to mention semiclassical regimes, irreducible representations, or other ingredients necessary to build a model. The geometry of Lorentzian 4-simplices emerges naturally from the vertices of any locally flat Lorentzian spin foam theory with $SU(2)$ edge invariance. 
What role do the details of the spin foam model play? The model has to introduce a scale $\hbar$ that distinguishes the semiclassical regime from the quantum one. In the semiclassical regime, the model provides an action that tells us how to glue 4-simplices together and how to recover (discrete) Einstein equations. 
In the EPRL model, $\hbar$ multiplies the spins associated with the areas of the geometry. Therefore, we identify the semiclassical regime with the large quantum numbers regime\footnote{To be more precise, the community recently agreed upon identifying the semiclassical regime as a double limit of large quantum numbers and refined 2-complexes.}.
We glue two 4-simplices together, identifying the shared tetrahedra up to the choice of frame. How to recover the Einstein equations in the theory is still unknown and is an object of active research. In this regime, the EPRL model also gives the edge closure conditions as saddle point equations and the action reduces to the area-angle Regge calculus area.  

\medskip

We believe this is the best starting point if we want to iterate and improve the EPRL spin foam model using a top-down approach starting from discrete general relativity. We realize this is an arduous and lengthy path and that this work is only the first step in this direction.

\medskip

We conclude with a few observations. The story we tell shows many connections with Effective Spin Foams\cite{Asante:2020iwm, Asante:2021zzh}. They abandon loop quantum gravity variables to work directly with flat Lorentzian 4-simplices, take the action of area-angle Regge calculus seriously, and study the gluing of 4-simplices with a term inspired by spin foams. In this work, we prove that the integral over $SL(2,\C)$ holonomies of the EPRL model is somehow equivalent to the sum over Lorentzian 4-simplices of the effective model. We leave to future work, a more in-depth study on how to translate their results into the holonomies language that talks directly to traditional covariant loop quantum gravity.

\medskip

There are numerous extensions of the EPRL model that includes 4-simplices with time-like boundary. Even if this work cannot apply directly to those models, local flatness is a fundamental ingredient also in that case. The parametrization \eqref{eq:wedgeholo} uses spinors that carry a representation of $SU(2)$, the subgroup of the Lorentz group stabilized by the normal of the space-like hypersurface. We can repeat our analysis by changing the parametrization of the wedge holonomies with objects that carry a $SU(1,1)$ representation. We expect to find comparable results also in that case.

\section{Acknowledgments}
This work was made possible through the support of the  FQXi  Grant  FQXi-RFP-1818 and of the ID\# 61466 grant from the John Templeton Foundation, as part of the ``The Quantum Information Structure of Spacetime (QISS)'' Project (\href{qiss.fr}{qiss.fr}). We want to thank Giorgio Sarno for the time spent studying \cite{Hellmann:2013gva} together. We thank Simone Speziale for countless discussions in the initial phase of this work. And finally, we thank Carlo Rovelli, Francesca Vidotto, and Hal Haggard for discussing the interpretation of the result.

%%%%%%%%%%%%%%%%%%%%%%%%%%%%%%%%%%%%%%%%%%%%%%%%%%%%%%%%%%%%%%%%%%%%%%%%%%%%%%%%%%%%%%%%%%

\begin{appendices}

%%%%%%%%%%%%%%%%%%%%%%%%%%%%%%%%%%%%%%%%%%%%%%%%%%%%%%%%%%%%%%%%%%%%%%%%%%%%%%%%%%%%%%%%%%

\section{Algebra with spinors}
\label{app:aspinor}
Given a unit spinor $\ket{z}$ such that $\bra{z} \vec{\sigma} \ket{z} = - \vec{n}$ we have that
\begin{equation}
\label{eq:projector1}
\ket{z}\bra{z} = \frac{\mathds{1} - \vec{n} \cdot \vec{\sigma}}{2} \ ,
\end{equation}
is a projector on the one dimensional subspace of $\C^2$ spanned by $\ket{z}$. The orthogonal projector is given by 
\begin{equation}
\label{eq:projector2}
\sket{z}\sbra{z} = \frac{\mathds{1} + \vec{n} \cdot \vec{\sigma}}{2} \ .
\end{equation}
The sum of the two orthogonal projectors provides a resolution of the identity of $\C^2$
\begin{equation}
	\ket{z}\bra{z} + \sket{z}\sbra{z} = \mathds{1} \ .
\end{equation}
The scalar products of two unit spinors and their duals are related by the followind identities
\begin{equation}
\label{eq:scalarproductprop}
	\sbrasket{z}{w} = \braket{w}{z}= \overline{\braket{z}{w}} \ , \qquad \sbraket{z}{w} = - \sbraket{w}{z} = -\overline{\brasket{z}{w}} \ . 
\end{equation}
Therefore, a direct calculation shows that
\begin{equation}
\label{eq:modulusbrasbra}
	\left|\braket{w}{z}\right|^2 = 1 - \left|\sbraket{w}{z}\right|^2 \ .
\end{equation}

\medskip

\noindent Given two unit spinors $\ket{z_{ab}}$ and $\ket{z_{ac}}$ representing two framed planes we compute 
\begin{equation}
	\label{eq:dihedralandmodulus}
	\left|  \braket{z_{ab}}{z_{ac}}\right|^2 = \braket{z_{ab}}{z_{ac}}\braket{z_{ac}}{z_{ab}} = \bra{z_{ab}}\frac{\mathds{1} - \vec{n}_{ac} \cdot \vec{\sigma}}{2}\ket{z_{ab}}= \frac{1 + \vec{n}_{ab} \cdot \vec{n}_{ac}}{2} = \frac{1 + \cos\phi^a_{bc}}{2} \ ,
\end{equation}
where $\cos\phi^a_{bc} = \vec{n}_{ab} \cdot \vec{n}_{ac}$ is the dihedral angle between the framed planes $\ket{z_{ab}}$ and $\ket{z_{ac}}$ orthogonal to $\vec{n}_{ab} $ and $\vec{n}_{ac}$. Using \eqref{eq:modulusbrasbra} we have 
\begin{equation}
	\label{eq:dihedralandmodulus2}
	\left|  \sbraket{z_{ab}}{z_{ac}}\right|^2 = \frac{1 - \cos\phi^a_{bc}}{2} \ .
\end{equation}
Combining \eqref{eq:dihedralandmodulus} and \eqref{eq:dihedralandmodulus2} and using basic trigonometry we also have 
\begin{equation}
\label{eq:sine}
2 \left|  \sbraket{z_{ab}}{z_{ac}} \braket{z_{ab}}{z_{ac}} \right| = 2 \sqrt{ \frac{1 - \cos\phi^a_{bc}}{2} } \sqrt{ \frac{1 + \cos\phi^a_{bc}}{2} } = \sin \phi^a_{bc} \ ,
\end{equation}
where we assumed by convention that $0\leq \phi^a_{bc} \leq \pi$.
%%%%%%%%%%%%%%%%%%%%%%%%%%%%%%%%%%%%%%%%%%%%%%%%%%%%%%%%%%%%%%%%%%%%%%%%%%%%%%%%%%%%%%%%%%

\section{Canonical form of the holonomies}
\label{app:diag}
Consider two spinors $\ket{z}$ and $\ket{w}$ representing the basis of two $\C^2$ spaces. If we interpret them as the source and target space of a linear map $g\in GL(2,\C)$ we can parametrize it using the projectors and complex coefficients
\begin{equation}
g = a \ket{w} \bra{z} + b \sket{w} \bra{z} + c \ket{w} \sbra{z} + d \sket{w} \sbra{z} \ .
\end{equation} 
A general $g \in SU(2)$ element can be parametrized by 
\begin{equation}
g = a \ket{w} \bra{z} + b \sket{w} \bra{z} -b^* \ket{w} \sbra{z} + a^* \sket{w} \sbra{z} \ ,
\end{equation} 
with the unit determinant condition $|a|^2 + |b|^2=1$. Moreover, we can always find a canonical basis for the source and target space for which 
\begin{equation}
g = e^{i\tfrac\phi2}\ket{w} \bra{z} + e^{-i\tfrac\phi2}\sket{w} \sbra{z} \ ,
\end{equation} 
where the phase $\phi$ is redundant and can be reabsorbed in the phase of any of the two spinors. We keep it anyway since it is convenient for our analysis. We can also prove that 
\begin{equation}
b = e^{\tfrac\eta2}\ket{w} \bra{w} + e^{-\tfrac\eta2}\sket{w} \sbra{w} \ ,
\end{equation} 
is an element of $SL(2,\C)$ and is a pure boost. Using the projectors \eqref{eq:projector1} and \eqref{eq:projector2} and denoting with $\vec{m}$ the normal of the framed plane $\ket{w}$
\begin{equation}
b = e^{\tfrac\eta2}\ket{w} \bra{w} + e^{-\tfrac\eta2}\sket{w} \sbra{w} = \frac{e^{\tfrac\eta2} + e^{-\tfrac\eta2}}{2} \mathds{1} - \frac{e^{\tfrac\eta2} - e^{-\tfrac\eta2}}{2} \vec{m} \cdot \vec{\sigma} = e^{-\eta \tfrac{\vec{m} \cdot \vec{\sigma}}{2}} \ ,
\end{equation} 
which is the canonical form of a pure boost with rapidity $\eta$ and axis $-\vec{m}$. It is clear, for arbitrary $\vec{m}$ ($\ket{w}$) and rapidity $\eta$ we obtain all the possible boosts. The canonical form \eqref{eq:gdiagonal} can be decomposed in a pure boost times an arbitrary rotation. Therefore, it represent the most general Lorentz transformation $g\in SL(2,\C)$
\begin{equation}
g = \left(e^{\frac{\Re\omega}{2}} \ket{w}\bra{w}+ e^{-\frac{\Re\omega}{2}} \sket{w}\sbra{w}\right) \left(e^{\frac{i\Im\omega}{2}} \ket{w}\bra{z}+ e^{-\frac{i\Im\omega}{2}} \sket{w}\sbra{z}\right) \ .
\end{equation}

%%%%%%%%%%%%%%%%%%%%%%%%%%%%%%%%%%%%%%%%%%%%%%%%%%%%%%%%%%%%%%%%%%%%%%%%%%%%%%%%%%%%%%%%%%

\section{Sperical cosine laws and local flatness}
\label{app:derivation_spherical}
This section derives the spherical cosine and sine laws from the local flatness conditions. The calculation is straightforward, although lengthy. We start from the local flatness condition \eqref{eq:3cycleflatness} where we isolate the wedge holonomy $g_{ab}$
\begin{equation}
\label{eq:localflat1}
g_{ab}^{-1} = g_{ca}g_{bc} \ .
\end{equation}
It is more convenient to work with scalar equations than matricial ones. We take the matrix elements
\begin{equation}
\begin{split}
\sbra{z_{ac}} g_{ab}^{-1} \ket{z_{bc}} = \sbra{z_{ac}} g_{ca}g_{bc} \ket{z_{bc}} \ , \\
\bra{z_{ac}} g_{ab}^{-1} \sket{z_{bc}} = \bra{z_{ac}} g_{ca}g_{bc} \sket{z_{bc}} \ ,
\end{split}
\end{equation}
with the intent of isolating the complex angle $\omega_{ab}$\footnote{The alternative projection $\bra{z_{ac}} \cdot \ket{z_{bc}}$ and $\sbra{z_{ac}} \cdot \sket{z_{bc}}$ leads to the same result with an equivalent path.}. Substituting the explicit form of the wedge holonomies \eqref{eq:wedgeholo} we get 
\begin{equation}
\begin{split}
e^{-\frac{\omega_{ab}}{2}}\sbraket{z_{ac}}{z_{ab}}\sbraket{z_{ba}}{z_{bc}}-e^{\frac{\omega_{ab}}{2}}\sbrasket{z_{ac}}{z_{ab}}\braket{z_{ba}}{z_{bc}}&=e^{\frac{\omega_{ca}+\omega_{bc}}{2}}\brasket{z_{ca}}{z_{cb}} \ , \\
e^{-\frac{\omega_{ab}}{2}}\braket{z_{ac}}{z_{ab}}\sbrasket{z_{ba}}{z_{bc}}-e^{\frac{\omega_{ab}}{2}}\brasket{z_{ac}}{z_{ab}}\brasket{z_{ba}}{z_{bc}}&=e^{-\frac{\omega_{ca}+\omega_{bc}}{2}}\sbraket{z_{ca}}{z_{cb}} \ .
\end{split}
\end{equation}
Taking the term-by-term product of these two equations, we can eliminate the complex angles $\omega_{ca}$, $\omega_{bc}$ and obtain
\begin{equation}
\begin{split}
e^{\omega_{ab}}\sbrasket{z_{ac}}{z_{ab}}\braket{z_{ba}}{z_{bc}}\brasket{z_{ac}}{z_{ab}}\brasket{z_{ba}}{z_{bc}} +
e^{-\omega_{ab}}\braket{z_{ac}}{z_{ab}}\sbrasket{z_{ba}}{z_{bc}}\sbraket{z_{ac}}{z_{ab}}\sbraket{z_{ba}}{z_{bc}}&\\
-\sbrasket{z_{ac}}{z_{ab}}\braket{z_{ba}}{z_{bc}}\braket{z_{ac}}{z_{ab}}\sbrasket{z_{ba}}{z_{bc}}
-\sbraket{z_{ac}}{z_{ab}}\sbraket{z_{ba}}{z_{bc}}\brasket{z_{ac}}{z_{ab}}\brasket{z_{ba}}{z_{bc}}&
=\brasket{z_{ca}}{z_{cb}}\sbraket{z_{ca}}{z_{cb}}
\end{split}
\end{equation}
To isolate the complex angle $\omega_{ab}$ it is convenient to manipulate the equality using the properties \eqref{eq:scalarproductprop} and the definition of the twist angle \eqref{eq:twistangle}
\begin{equation}
\cosh(\omega_{ab} + i \xi^c_{ab})
=\frac{-|\brasket{z_{ca}}{z_{cb}}|^{2}+|\braket{z_{ab}}{z_{ac}}|^{2}|\braket{z_{ba}}{z_{bc}}|^{2}+|\brasket{z_{ac}}{z_{ab}}|^{2}|\brasket{z_{ba}}{z_{bc}}|^{2}}{|\braket{z_{ac}}{z_{ab}}\brasket{z_{ac}}{z_{ab}}\braket{z_{ba}}{z_{bc}}\brasket{z_{ba}}{z_{bc}}|} \ .
\end{equation}
Using the formulas to relate the absolute values of the spinor scalar products to the dihedral angles between framed planes \eqref{eq:dihedralandmodulus}, \eqref{eq:dihedralandmodulus2} and \eqref{eq:sine} we finally obtain \eqref{eq:cosine_equations}
\begin{equation}
\cosh(\omega_{ab} + i \xi^c_{ab})
=\frac{\cos\phi_{ab}^{c}+\cos\phi_{bc}^{a}\cos\phi_{ac}^{b}}{\sin\phi_{bc}^{a}\sin\phi_{ac}^{b}} \ .
\end{equation}
To find the equivalent equations for the complex angles $\omega_{ca}$ and $\omega_{bc}$ we repeat the calculation isolating $g_{ca}$ and $g_{bc}$ at the fist step \eqref{eq:localflat1}.

\medskip

The derivation of \eqref{eq:sine_equations} is more complicated but involves only basic algebra and the spinorial calculus properties we introduced in Appendix~\ref{app:aspinor}\footnote{We suspect there is a more straightforward derivation of these equations.}. We will sketch the main steps and omit the majority of the algebra. We start with a particular combination of projections of the matricial equation \eqref{eq:localflat1}
\begin{align}
\label{eq:fist_sine_step}
\sbra{z_{ac}} g_{ab}^{-1} \ket{z_{bc}}\bra{z_{ac}} g_{ab}^{-1} \ket{z_{bc}} &= \sbra{z_{ac}} g_{ca}g_{bc} \ket{z_{bc}}\bra{z_{ac}} g_{ca}g_{bc} \ket{z_{bc}} \ , \\
\bra{z_{ac}} g_{ab}^{-1} \sket{z_{bc}}\sbra{z_{ac}} g_{ab}^{-1} \sket{z_{bc}} &= \bra{z_{ac}} g_{ca}g_{bc} \sket{z_{bc}}\sbra{z_{ac}} g_{ca}g_{bc} \sket{z_{bc}} \ . 
\end{align}
Then we substitute the explicit expression for the wedge holonomies \eqref{eq:wedgeholo}, we divide the first equation by $\braket{z_{ba}}{z_{bc}}\sbraket{z_{ba}}{z_{bc}}$, the second by $\brasket{z_{ba}}{z_{bc}}\sbrasket{z_{ba}}{z_{bc}}$ and we subtract them. The result, after a few simplifications, is
\begin{equation}
e^{\omega_{ab}}\frac{\brasket{z_{ac}}{z_{ab}}\sbrasket{z_{ac}}{z_{ab}}}{\sbraket{z_{ba}}{z_{bc}}\braket{z_{bc}}{z_{ba}}}
-
e^{-\omega_{ab}}\frac{\overline{\brasket{z_{ac}}{z_{ab}}}\overline{\sbrasket{z_{ac}}{z_{ab}}}}{\overline{\sbraket{z_{ba}}{z_{bc}}}\overline{\braket{z_{bc}}{z_{ba}}}}
= 
e^{\omega_{bc}}\frac{\braket{z_{cb}}{z_{ca}}\brasket{z_{ca}}{z_{cb}}}{\braket{z_{ba}}{z_{bc}}\sbraket{z_{ba}}{z_{bc}}}
-
e^{-\omega_{bc}}\frac{\overline{\braket{z_{cb}}{z_{ca}}}\overline{\brasket{z_{ca}}{z_{cb}}}}{\overline{\braket{z_{ba}}{z_{bc}}}\overline{\sbraket{z_{ba}}{z_{bc}}}} \ .
\end{equation}
If we extract the phases from the scalar product, we can recognize the twist angles and collect the remaining absolute value
\begin{equation}
2\sinh(\omega_{ab} + i \xi^c_{ab})\frac{|\brasket{z_{ac}}{z_{ab}}\braket{z_{ac}}{z_{ab}}|}{|\brasket{z_{ba}}{z_{bc}}\braket{z_{bc}}{z_{ba}}|}
= 
2\sinh(\omega_{bc}+ i \xi^a_{bc})\frac{|\braket{z_{cb}}{z_{ca}}\brasket{z_{ca}}{z_{cb}}|}{|\braket{z_{ba}}{z_{bc}}\sbraket{z_{ba}}{z_{bc}}|} \ .
\end{equation} 
In terms of the dihedral angles between framed planes \eqref{eq:dihedralandmodulus}, \eqref{eq:dihedralandmodulus2} and \eqref{eq:sine} we obtain directly \eqref{eq:sine_equations}
\begin{equation}
\sinh(\omega_{ab} + i \xi^c_{ab})\sin \phi^a_{bc}
= 
\sinh(\omega_{bc}+ i \xi^a_{bc})\sin \phi^c_{ab} \ .
\end{equation} 
If in \eqref{eq:fist_sine_step} we start from the product of the projections $\sbra{z_{ac}} \cdot \ket{z_{bc}}\sbra{z_{ac}} \cdot \sket{z_{bc}}$ and the complementary one we obtain the analog equation relating the complex angles $\omega_{ab}$ and $\omega_{ca}$.

\section{Singular support of $F_{EPRL}$}
\label{app:singularsupport}
We start from the expression \eqref{eq:WEPRL}, and we explicit the form of the $\gamma$-simple unitary irreducible representations. We find 
\begin{equation}
\label{eq:WEPRLexpand}
\begin{split}
F_{EPRL}(g_{ab},\tilde{g}_{ba})=\sum_{j_{ab}} (2j_{ab}+1)& \int (2j_{ab}+1) \d \zeta_{ab} \int (2j_{ab}+1) \d \zeta_{ba} \int (2j_{ab}+1) \d \tilde\zeta_{ab} \int (2j_{ab}+1) \d \tilde\zeta_{ba} \\
&
(2j_{ab}+1)\int \d z_{ab}
\frac{\sbraket{\zeta_{ba}}{z_{ab}}^{2j_{ab}}\bra{z_{ab}}g_{ab}\ket{\zeta_{ab}}^{2j_{ab}}}{\Vert z_{ab}\Vert ^{2j_{ab}+2i\gamma j_{ab}+2} \Vert g_{ab}^{\dagger}z_{ab}\Vert ^{2j_{ab}-2i\gamma j_{ab}+2}}
\brasket{\zeta_{ab}}{\tilde\zeta_{ab}}^{2 j_{ab}}\\
&(2j_{ab}+1)\int \d \tilde z_{ba}
\frac{\sbraket{\tilde\zeta_{ab}}{\tilde z_{ba}}^{2j_{ab}}\bra{\tilde z_{ba}}\tilde g_{ba}\ket{\tilde\zeta_{ba}}^{2j_{ab}}}{\Vert \tilde z_{ba}\Vert ^{2j_{ab}+2i\gamma j_{ab}+2} \Vert \tilde g_{ba}^{\dagger}\tilde z_{ba}\Vert ^{2j_{ab}-2i\gamma j_{ab}+2}}
\brasket{\tilde\zeta_{ba}}{\zeta_{ba}}^{2 j_{ab}} \ .\\
\end{split}
\end{equation}
We group the integrations over the various spinors under a single symbol $\d \Omega$, and we collect the integer $2j_{ab}$ from the exponent 
\begin{equation}
\label{eq:WEPRLreason}
F_{EPRL}(g_{ab},\tilde{g}_{ba})= \sum_{j_{ab}} (2j_{ab}+1)^7 \int d\Omega \left( f_{EPRL}(g_{ab},\tilde{g}_{ba}) \right)^{j_{ab}} \ ,
\end{equation}
where we denoted with 
\begin{equation}
\label{eq:wEPRL}
f_{EPRL}(g_{ab},\tilde{g}_{ba}) =  \frac{\sbraket{\zeta_{ba}}{z_{ab}}\bra{z_{ab}}g_{ab}\ket{\zeta_{ab}}}{\Vert z_{ab}\Vert^{1+i\gamma}\Vert g_{ab}^{\dagger}z_{ab}\Vert^{1-i\gamma}}\brasket{\zeta_{ab}}{\tilde{\zeta}_{ab}}\frac{\sbraket{\tilde{\zeta}_{ab}}{\tilde{z}_{ba}}\bra{\tilde{z}_{ba}}\tilde{g}_{ba}\ket{\tilde{\zeta}_{ba}}}{\Vert\tilde{z}_{ba}\Vert^{1+i\gamma}\Vert\tilde{g}_{ba}^{\dagger}\tilde{z}_{ba}\Vert^{1-i\gamma}}\brasket{\tilde{\zeta}_{ba}}{\zeta_{ba}} \ .
\end{equation}
We recast the sum over half-integer spins in terms of integers since $2j_{ab}\in \mathbb{N}$. The function $F_{EPRL}(g_{ab},\tilde{g}_{ba})$ has the form of (the seventh derivative of) a geometric series that is defined in the whole complex plane but $1$ where it is singular. Therefore we find the singular support of \eqref{eq:WEPRLreason} for values of the holonomies such that $w_{EPRL}(g_{ab},\tilde{g}_{ba})=1$ or 
\begin{equation}
\label{eq:condition}
\frac{\sbraket{\zeta_{ba}}{z_{ab}}\bra{z_{ab}}g_{ab}\ket{\zeta_{ab}}}{\Vert z_{ab}\Vert^{1+i\gamma}\Vert g_{ab}^{\dagger}z_{ab}\Vert^{1-i\gamma}}\brasket{\zeta_{ab}}{\tilde{\zeta}_{ab}}\frac{\sbraket{\tilde{\zeta}_{ab}}{\tilde{z}_{ba}}\bra{\tilde{z}_{ba}}\tilde{g}_{ba}\ket{\tilde{\zeta}_{ba}}}{\Vert\tilde{z}_{ba}\Vert^{1+i\gamma}\Vert\tilde{g}_{ba}^{\dagger}\tilde{z}_{ba}\Vert^{1-i\gamma}}\brasket{\tilde{\zeta}_{ba}}{\zeta_{ba}} =1 \ .
\end{equation}
Note that we are integrating all the spinors. In general, there are no holonomies such that \eqref{eq:condition} is satisfied. Taking the absolute value of the left-hand side of the \eqref{eq:condition} and applying the Cauchy-Schwarz inequality to all the scalar products at the numerator 
\begin{equation}
\label{eq:condition2}
\left|f_{EPRL}(g_{ab},\tilde{g}_{ba})\right|=
\left|\frac{\sbraket{\zeta_{ba}}{z_{ab}}\bra{z_{ab}}g_{ab}\ket{\zeta_{ab}}}{\Vert z_{ab}\Vert \Vert g_{ab}^{\dagger}z_{ab}\Vert}\brasket{\zeta_{ab}}{\tilde{\zeta}_{ab}}\frac{\sbraket{\tilde{\zeta}_{ab}}{\tilde{z}_{ba}}\bra{\tilde{z}_{ba}}\tilde{g}_{ba}\ket{\tilde{\zeta}_{ba}}}{\Vert\tilde{z}_{ba}\Vert\Vert\tilde{g}_{ba}^{\dagger}\tilde{z}_{ba}\Vert}\brasket{\tilde{\zeta}_{ba}}{\zeta_{ba}} \right| \leq 1 \ .
\end{equation}
The inequality is saturated if all the vectors in the scalar product are proportional, in particular, if for some real parameters $\rho_i\in \R$ and $\beta_i\in\R$\begin{align}
\label{eq:condition3}
\ket{z_{ab}}		     &=e^{i\beta_{1}}\sket{\zeta_{ba}}&
g_{ab}\ket{\zeta_{ab}}   &=e^{\rho_{2}+i\beta_{2}}\ket{z_{ab}}&
\sket{\tilde{\zeta}_{ab}}&=e^{i\beta_{3}}\ket{\zeta_{ab}}\\
\ket{\tilde{z}_{ba}}	 &=e^{i\beta_{4}}\sket{\tilde{\zeta}_{ab}}&
\tilde{g}_{ba}\ket{\tilde{\zeta}_{ba}} &= e^{\rho_{5}+i\beta_{5}}\ket{\tilde{z}_{ba}} &
\sket{\zeta_{ba}}&=e^{i\beta_{6}}\ket{\tilde{\zeta}_{ba}}  \nonumber \ .
\end{align}
Combining the equations \eqref{eq:condition3} we find an eigenvalue equations for the face holonomy
\begin{equation}
\tilde{g}_{ba} g_{ab} \ket{\zeta_{ab}} = e^{\frac{\omega_f}{2}} \ket{\zeta_{ab}} \ ,
\end{equation}
where $\tfrac{\omega_f}{2} = \rho_2 + \rho_5 + i (\beta_1 + \beta_2 + \beta_3 + \beta_4 + \beta_5 + \beta_6)$. From the Cauchy-Schwarz inequality we also get the extra equations
\begin{align}
\label{eq:condition4}
\ket{\zeta_{ab}}   &=e^{\rho_{7}-i\beta_{7}}(g_{ab}^\dagger)^{-1} \ket{z_{ab}}&
\ket{\tilde{\zeta}_{ba}} &= e^{\rho_{8}-i\beta_{8}}(\tilde{g}_{ba}^\dagger)^{-1}\ket{\tilde{z}_{ba}} \ ,
\end{align}
or equivalently
\begin{align}
\label{eq:condition5}
g_{ab}\sket{\zeta_{ab}}   &=e^{\rho_{7}+i\beta_{7}} \sket{z_{ab}}&
\tilde{g}_{ba}\ket{\tilde{\zeta}_{ba}} &= e^{\rho_{8}+i\beta_{8}} \ket{\tilde{z}_{ba}} \ .
\end{align}
From these equations we find that also $\sket{\zeta_{ab}}$ is an eigenvector of the face holonomy 
\begin{equation}
\tilde{g}_{ba} g_{ab} \sket{\zeta_{ab}} = e^{\frac{\tilde \omega_f}{2}} \sket{\zeta_{ab}} \ .
\end{equation}
From the orthonormality of the eigenvectors and $\det \tilde{g}_{ba} g_{ab} =1$ we conclude that $\tilde \omega_f = - \omega_f$. 
We insert the equations \eqref{eq:condition3} and \eqref{eq:condition4} into \eqref{eq:condition} to derive a condition on the real and imaginary part of the face complex angle 
\begin{equation}
\label{eq:faceWFS}
\frac{\gamma \Re \omega_f + \Im \omega_f}{2} = 0 \ \mathrm{mod} \ 2\pi \qquad \text{ or }\qquad \gamma \Re \omega_f + \Im \omega_f = 4 k \pi \qquad k\in \mathbb{Z} \ .
\end{equation}
Since we are integrating over $\ket{\zeta_{ab}}$ we can conclude that $\tilde{g}_{ba} g_{ab}$ can be an arbitrary 4-screw with the relation \eqref{eq:faceWFS} between rapidity and rotation angle.
\begin{equation}
\mathrm{Sing} (F_{EPRL}(g_{ab},\tilde{g}_{ba})) = \left\lbrace g_{ab},\tilde{g}_{ba}\ \vert \ \tilde{g}_{ba} g_{ab} = 
e^{\frac{\omega_{f}}{2}} \ket{\zeta}\bra{\zeta} + e^{-\frac{\omega_{f}}{2}} \sket{\zeta}\sbra{\zeta} \ \text{with} \ \gamma \Re \omega_f + \Im \omega_f = 0 \ \mathrm{mod} \ 4\pi \right\rbrace \ .
\end{equation}
The calculation for a general face with any amount of wedges (vertices) is the same. For example, the singular support of the EPRL function associated with a face with  $1, \ldots, n$ wedges is 
\begin{equation}
\mathrm{Sing} (F_{EPRL}(g_{1},\, \ldots,\, g_{n})) = \left\lbrace (g_{1},\, \ldots,\, g_{n})\ \vert \ g_f = 
e^{\frac{\omega_{f}}{2}} \ket{\zeta}\bra{\zeta} + e^{-\frac{\omega_{f}}{2}} \sket{\zeta}\sbra{\zeta} \ \text{with} \ \gamma \Re \omega_f + \Im \omega_f = 0 \ \mathrm{mod} \ 4\pi \right\rbrace \ ,
\end{equation}
Where $g_f = g_n \cdots g_1$ is just the (ordered) product of all the wedge holonomies belonging to that face. 
%%%%%%%%%%%%%%%%%%%%%%%%%%%%%%%%%%%%%%%%%%%%%%%%%%%%%%%%%%%%%%%%%%%%%%%%%%%%%%%%%%%%%%%%%%

\end{appendices}

%%%%%%%%%%%%%%%%%%%%%%%%%%%%%%%%%%%%%%%%%%%%%%%%%%%%%%%%%%%%%%%%%%%%%%%%%%%%%%%%%%%%%%%%%%

\end{document}